\documentclass[preprint,12pt]{article}

\usepackage{amssymb}
\usepackage{alltt}
\usepackage{epsfig}
\begin{document}

\begin{flushright}
CP3-Origins-2011-27\\
DIAS-2011-14
\end{flushright}

\vspace{1 cm}

\begin{center}
{\Large{\bf  The Einstein-Maxwell system, Ward identities, and the Vilkovisky construction}}
\end{center}

\begin{center}
{\large{ N. K. Nielsen}}
\end{center}

\begin{center}
Department of Physics and Chemistry,\\
University of Southern Denmark, \\
Odense, Denmark
\end{center}

\begin{abstract}
The gauge fixing dependence of the one-loop effective action of quantum gravity in the proper-time representation is investigated for a space of arbitrary curvature, and the investigation  is extended to Maxwell-Einstein theory.  The construction of Vilkovisky and DeWitt for removal of this depence is then considered in general gauges, and it is shown that nontrivial criteria arising from a Ward identity of the theory must be obeyed by the regularization scheme, if the construction is to remove the gauge dependence of quadratic and quartic divergences.  The results apply also to non-Abelian gauge theories; they  are used to address the question of gauge dependence of asymptotic freedom arising through internal graviton lines at one-loop order as suggested by Robinson and Wilczek. 
 \noindent {\em PACS  numbers 04.62.+v, 11.10.Gh, 11.15.-q}.\\
\end{abstract}

%\end{frontmatter}

\DeclareGraphicsRule{.tif}{png}{.png}{`convert #1 `dirname #1`/`basename #1 .tif`.png}
\setlength{\textwidth}{14.5cm}
\setlength{\textheight}{23.5cm}
\setlength{\topmargin}{-2cm}
\setlength{\evensidemargin}{0.11cm}
\setlength{\oddsidemargin}{0.11cm}

\newpage

\section{Introduction}

In an influential   paper, Robinson and Wilczek \cite{Robinson} suggested the possibility of asymptotic freedom arising in a gauge theory, considered an effective field theory in the sense of Weinberg \cite{Weinberg}, Vilkovisky \cite{VilkoviskyIII} and Donoghue \cite{Donoghue}, through the quadratic divergences of one-loop Feynman integrals involving internal graviton lines.  It was subsequently pointed out by Pietrykowski \cite{Pietrykowski}  that the effect is gauge dependent and that it vanishes in a class of gauges. This was later confirmed in \cite{Rodigast}. The topic has recently generated much active interest and controversy in the litterature \cite{Ellis}.

The results in \cite{Robinson}-\cite{Ellis} were based on flat-space calculations, which means that the background gravitational field is not a solution of the Einstein equation, and this in its turn leads to a gauge-dependent result, since off-shell quantities in a quantum field theory depend on the gauge. A gauge-independent construction of the off-shell effective action was developed by Vilkovisky \cite{Vilkovisky} and extended by DeWitt \cite {DeWittI} (we shall for brevity refer to this method below as the  Vilkovisky construction). Their method was recently applied in connection with the problem treated in \cite{Robinson}-\cite{Ellis} by Toms \cite{Toms}, \cite{TomsII} using the Schwinger-DeWitt proper-time representation of the effective action \cite{Schwinger}, \cite{DeWittII}, and by He, Wang and Xianyu \cite{He} and  Tang and Wu \cite{Tang}, using momentum space integration. They  all obtain different numerical results.

The scope of the present investigation is threefold:
\begin{itemize}
\item The problem of \cite{Robinson} is considered in a space with  arbitrary curvature by means of the Schwinger-De Witt proper-time representation of the effective action of the gravitational field in arbitrary gauges; thus the Einstein equation for the background metric may be applied, formally eliminating the gauge dependence of the effective action, and also the consequences in general of having a background metric that is not a solution of the Einstein equation can be found. 
\item  The Vilkovisky construction  is investigated in detail on the one-loop level in general gauges rather than the so-called Landau-DeWitt gauge \cite{Fradkin} to which  \cite{Toms}, \cite{TomsII}, \cite{He} and   \cite{Tang} were restricted. It is found that gauge independence of the one-loop effective action  is a consequence of a certain  Ward identity, and the Vilkovisky construction can thus only be applied in connection with regularization schemes  where this Ward identity is not violated.  
\item This method is then applied to the Maxwell-Einstein system, and it is found that its one-loop effective action is made gauge-invariant off-shell at second order in the gravitational coupling $\kappa$ by the Vilkovisky construction of pure quantum gravity,  whereas the version of the Landau-DeWitt gauge used in \cite{Toms}, \cite{TomsII}, \cite{He} and \cite{Tang} is relevant  for the off-shell effective action at fourth order in  $\kappa$. 
\end{itemize}
Because the topic has generated so much controversy a rather detailed exposition has been used.  Only coupled Maxwell-Einstein fields are considered, but the conclusions carry over almost verbatim to non-Abelian gauge fields coupled to gravity.

The layout of this article is the following: In sec. 2 we  consider a general gauge theory and carry out  the Vilkovisky construction in the one-loop approximation,  showing how  a general Ward identity is formally valid and implies through  a partial cancellation of the gauge field and ghost contribution  that the effective action is  independent of the gauge condition.

In sec. 3 we consider the one-loop effective action of pure quantum gravity in an arbitrary background metric and in a class of gauges more general than the Feynman gauge, showing that the Schwinger-DeWitt proper time representation can be used also in this case, and that the gravitational heat kernel obeys a Ward identity that determines the gauge dependence of the effective action in the case where the background metric is not a solution of the Einstein equation.  The Vilkovisky construction is also carried out in this case and shown not to eliminate the gauge dependence of all quadratically divergent terms. It is found that the gauge-dependent part of the effective action contains in the proper-time representation, apart from a term involving the background field Einstein equation and thus vanishing on-shell, also a quadratic divergence involving the scalar curvature  that cannot be eliminated by the Vilkovisky construction, which applies on the non-regularized level.   

 In sec. 4 the Maxwell-Einstein system is  considered, and it is found that the considerations on the gauge dependence of the effective action  obtained in sec. 3 carry over to this case also,  with the replacement ${\cal G} ^{\mu \nu}\rightarrow  {\cal G} ^{\mu \nu}-{\cal T} ^{\mu \nu}$, where ${\cal G} ^{\mu \nu}$ is the Einstein tensor and ${\cal T}^{\mu \nu}$ the energy-momentum tensor of the background gauge field. Also a class of generalized gauges is introduced involving the background gauge field and field strength, following \cite{Toms}, \cite{He} and \cite{Tang}, and it is proven that the contribution at second order in the gravitational coupling $\kappa$ formally vanishes by the Ward identities of the theory when the ghost contributions are taken into account, but that this is upset by quadratic divergences with proper time regularization.    We also  apply the Vilkovisky construction to this case.  As in the case of pure gravity not all gauge-parameter dependent quadratic divergences with a proper-time cut-off  are removed by the Vilkovisky construction. 
 
 Finally sec. 5 contains evaluation of the effective action  by momentum-space integration and application of the Vilkovisky construction in general gauges using this method of evaluation. Appendix A gives technical details on proper-time regularization while Appendix B contains the outline of the Vilkovisky construction of Maxwell-Einstein theory in next-lowest order in $\kappa$.

The following conventions have been used: The metric in Minkowski space is $\eta _{\mu \nu}=(-+++)$, and the sign of the Riemann tensor is chosen such that the Hilbert action is:
\begin{equation}
 S_{H}=\frac{1}{\kappa^2}\int d^4x\sqrt{-g}R
 \label{Hilbert}
 \end{equation}
 where $\kappa=\sqrt{8\pi G}$ is the gravitational coupling constant, with $G$ denoting Newton's constant.
 
 \
 \

 \section{General gauge theory}

The effective action $\Gamma [\phi]$ of a field theory with classical action $S[\varphi]$ is given by the path integral through:
\begin{equation}
e^{i\Gamma[\phi]}=\int [{\cal D}\varphi]\exp(i \Gamma[\phi]_{,k}(\phi^k- \varphi ^k)+S[\varphi]).
\label{limfjorden}
\end{equation}
We use the condensed notation of DeWitt \cite{DeWittII}, where the label of the background field $\phi$ and the integration variable $\varphi$ indicates both space-time variable, tensor indices and group indices. 
  From (\ref{limfjorden}) follows in the one-loop approximation:
 \begin{eqnarray}&&
 \Gamma[\phi]  \simeq S[\phi]  -\frac i2 {\rm Tr} \log \Delta[\phi].
 \label{snefoged}
 \end{eqnarray}
 The propagator $\Delta [\phi]^{ik}$ is defined by:
 \begin{equation}
 \Delta[\phi] ^{ik}S[\phi],_{kl}=\delta ^i\hspace{0.1 mm}_l.
 \label{skovfoged}
 \end{equation}
Vilkovisky \cite{Vilkovisky} and De Witt \cite{DeWittI} use in (\ref{limfjorden}) instead of the difference between the background field $\phi$ and the integration variable $\varphi$ the geodesic interval in field space, given a suitable metric tensor in field space, and this leads in (\ref{limfjorden}) to the replacement:
\begin{equation}
\phi^k-\varphi^k
\rightarrow \phi ^k-\varphi ^k-\frac 12 \Gamma ^k\hspace{0.1 mm}_{lm}[\phi](\phi ^l-\varphi ^l)(\phi ^m-\varphi ^m)+\dots
\label{violasilas}
\end{equation}
with  $\Gamma ^k\hspace{0.1 mm}_{lm}$ components of a connection in field space. Then  (\ref{snefoged}) becomes:
  \begin{equation}
 \Gamma[\phi]  \simeq S[\phi]
  -\frac i2 {\rm Tr} \log  \tilde \Delta[\phi]
 \label{protector}
 \end{equation}
 with:
\begin{equation}
\tilde  \Delta[\phi_0] ^{jk}(S[\phi_0],_{kl}-S[\phi_0]_{,m}\Gamma ^m\hspace{0.1 mm}_{kl}[\phi_0]))=\delta ^j\hspace{0.1 mm}_l
 \label{ridefoged}
 \end{equation}
 cp. (\ref{skovfoged}).  The general expression (\ref{protector}) can be written as a series in $S[\phi]_{,i}$ with the two first terms:
  \begin{equation}
 \Gamma[\phi]  \simeq S[\phi]
  -\frac i2 {\rm Tr} \log   \Delta[\phi]
   -\frac i2S[\phi]_{,m}\Gamma ^m\hspace{0.1 mm}_{kl}[\phi]\Delta [\phi ]^{lk}+\cdots .
 \label{ramses}
 \end{equation}
 This procedure corresponds to treating the connection term as an interaction term of the Lagrangian \cite{Toms}, \cite{TomsII}, \cite{He}, \cite{Tang}, \cite{TomsIII}.

 The effective action (\ref{protector}) can be considered an infinite series in $S_{,i}[\phi]$, where the  invariances achieved through the Vilkovisky construction (reparametrization invariance and also gauge fixing independence for a gauge theory) arise by conspiracy between neighboring terms. Truncating the series one will only have these invariances at a certain order.
 Since (\ref{ramses}) only contains $S_{,i}[\phi]$ in first order all terms of second and higher order   should be disregarded everywhere.

 Gauge transformations are:
 \begin{equation}
 \delta \phi^i=R^i\hspace{0.1 mm}_\alpha[\phi]\delta \lambda ^\alpha
 \end{equation}
 with $\delta \lambda ^\alpha$ an infinitesimal parameter, and where the gauge invariance of the classical action is expressed by:
 \begin{equation}
 S[\phi]_{,i}R^i\hspace{0.1 mm}_\alpha[\phi]=0.
 \label{miacis}
 \end{equation}
 From (\ref{miacis}) follows:
 \begin{equation}
R^i\hspace{0.1 mm}_\alpha[\phi] S[\phi]_{,ij}+ S[\phi]_{,i}R^i\hspace{0.1 mm}_\alpha[\phi]_{,j}=0
 \label{unfuwain}
 \end{equation}
 so $ S[\phi]_{,ij}$ is degenerate and not invertible on the mass shell, where $S[\phi]_{,i}=0$.
  The gauge transformation generators $ R^i\hspace{0.1 mm}_{\alpha}[\phi]$ fulfill the structure relations:
 \begin{equation}
 R^i\hspace{0.1 mm}_{\alpha, j}R^j\hspace{0.1 mm}_\beta- R^i\hspace{0.1 mm}_{\beta, j}R^j\hspace{0.1 mm}_\alpha=c^\gamma \hspace{0.1 mm}_{\alpha \beta}R^i\hspace{0.1 mm}_\gamma
 \label{derby}
 \end{equation}
where $c^\gamma \hspace{0.1 mm}_{\alpha \beta}$ are generalized structure constants.

 Gauge conditions are $\chi^\alpha[\phi]$ that are taken linear in $\phi$, and the  degeneracy  of  $ S[\phi]_{,ij}$ is lifted by  including in $S[\phi]_{,ij}$ the expression:
 \begin{equation}
\chi ^\alpha\hspace{0.1 mm}_{, i }c_{\alpha \beta}\chi ^\beta\hspace{0.1 mm} _{, j}
 \label{everttaube}
 \end{equation}
 with $c_{\alpha \beta}$ a constant, symmetric matrix, and  (\ref{protector}) is in a gauge theory replaced by:
  \begin{equation}
 \Gamma[\phi]\simeq  S[\phi]  -\frac i2 {\rm Tr} \log   \Delta[\phi]+i{\rm Tr}\log Q^{-1}-\frac i2S[\phi]_{,i}\tilde\Gamma ^i\hspace{0.1 mm}_{kl}[\phi]\Delta [\phi ]^{lk}
 \label{gamewarden}
 \end{equation}
with terms containing more than one power of $S[\phi]_{,i}$ disregarded and  with:
\begin{equation}
Q^\alpha \hspace{0.1 mm}_\beta=\chi^\alpha \hspace{0.1 mm}_{,i}R^i\hspace{0.1 mm}_\beta, \ {\rm det}Q^\alpha \hspace{0.1 mm}_\beta\neq 0
\label{gunnarson}
\end{equation}
where the new connection coefficients $\tilde\Gamma ^i\hspace{0.1 mm}_{kl}$ were constructed by Vilkovisky \cite{Vilkovisky}. After gauge-breaking  one gets by  inclusion of the expression (\ref{everttaube})   in $S[\phi]_{,ij}$ instead of (\ref{unfuwain}):
\begin{equation}
R^i\hspace{0.1 mm}_\alpha[\phi]S[\phi]_{,ij}+ S[\phi]_{,i}R^i\hspace{0.1 mm}_\alpha[\phi]_{,j}=Q^\beta \hspace{0.1 mm}_\alpha c_{\beta \gamma}\chi^\gamma \hspace{0.1 mm}_{,j}.
\label{ajdal}
\end{equation}
Multiplying this relation by  $(Q^{-1})^{\alpha}\hspace{0.1 mm}_\gamma\Delta[\phi]^{jk}$ one obtains the Ward identity \cite{Barvinsky}:
\begin{eqnarray}&&
c_{\alpha \beta }\chi ^\beta \hspace{0.1 mm}_{,j}\Delta [\phi]^{jk}=(Q^{-1})^{\beta}\hspace{0.1 mm}_\alpha(R^k\hspace{0.1 mm}_\beta[\phi]+ S[\phi]_{,i}R^i\hspace{0.1 mm}_{\beta,j}\Delta^{jk}[\phi]).
\label{royalacademy}
\end{eqnarray}
As mentioned above and as explained in more detail in the following, the gauge fixing independence of the effective action arises through conspiracy of neighboring terms of (\ref{protector}) when considered an infinite series in $S[\phi]_{,i}$ in the series expansion of the effective action. The Ward Identity (\ref{royalacademy}) is necessary for proving this gauge fixing indepence; notice that it contains terms of both zeroth and first order in  $S[\phi]_{,i}$.

 We introduce  the operator:
\begin{equation}
N_{\alpha \beta}=R^i\hspace{0.1 mm}_\alpha \gamma_{ij}R^j\hspace{0.1 mm}_\beta
\label{eliezer}
\end{equation}
with the inverse $N^{\alpha\beta}$.  Here $\gamma_{mn}$ is the  metric in field space, with the inverse metric $\gamma ^{mn}$. We also define the projection operators:
\begin{equation}
\Pi_{mn}=\gamma _{mn}-\gamma_{mi}R^i\hspace{0.1 mm}_\alpha N^{\alpha \beta}R^k\hspace{0.1 mm}_\beta\gamma_{kn}
\label{stamherren}
\end{equation}
with:
\begin{equation}
R^m \hspace{0.1 mm}_\alpha \Pi_{mn}=0.
\label{slotsherren}
\end{equation}

The connection in field space after gauge fixing $\tilde \Gamma ^r\hspace{0.1 mm}_{mn}$ is according to Vilkovisky \cite{Vilkovisky}, \cite{DeWittIII} given by:
\begin{eqnarray}&&
\tilde \Gamma ^r\hspace{0.1 mm}_{mn}
\simeq  \Gamma ^r\hspace{0.1 mm}_{mn} -\gamma_{mk}R^k\hspace{0.1 mm}_\alpha N^{\alpha \beta}{\cal D}_n R^r\hspace{0.1 mm}_{\beta}-\gamma_{nk}R^k\hspace{0.1 mm}_\alpha N^{\alpha \beta}{\cal D}_m R^r\hspace{0.1 mm}_{\beta}
 \nonumber\\&&
+\frac 12  \gamma_{mi}R^i\hspace{0.1 mm}_\alpha N^{\alpha \gamma}(R^j\hspace{0.1 mm}_\delta {\cal D}_jR^r\hspace{0.1 mm}_{\gamma}+R^j\hspace{0.1 mm}_{\gamma}{\cal D}_jR^r\hspace{0.1 mm}_{\delta}) N^{\delta \beta}R^k\hspace{0.1 mm}_\beta\gamma_{kn}
\label{fram}
\end{eqnarray}
with the covariant derivative defined through:
\begin{eqnarray}&&
{\cal D}_nR^r\hspace{0.1 mm}_{\alpha}=R^r\hspace{0.1 mm}_{\alpha,n}+\Gamma^r\hspace{0.1 mm}_{mn}R^m\hspace{0.1 mm}_\alpha.
\label{vullerslev}
\end{eqnarray}
and with $\Gamma ^r\hspace{0.1 mm}_{mn}$ the Christoffel connection components in field space before gauge fixing. The following  equivalent form of (\ref{fram}) turns out to be convenient:
\begin{eqnarray}&&
\tilde \Gamma ^r\hspace{0.1 mm}_{mn}\simeq \Gamma ^r\hspace{0.1 mm}_{ij}\Pi^i\hspace{0.1 mm}_m\Pi^j\hspace{0.1 mm}_n-\frac 12R^r\hspace{0.1 mm}_{\alpha, n} N^{\alpha \beta}R^k\hspace{0.1 mm}_{\beta}\gamma_{km}-\frac 12R^r\hspace{0.1 mm}_{\alpha, m} N^{\alpha \beta}R^k\hspace{0.1 mm}_{\beta}\gamma_{kn}
\nonumber\\&&
-\frac 12\Pi^s\hspace{0.1 mm}_mR^r\hspace{0.1 mm}_{\alpha, s}N^{\alpha \beta}R^k\hspace{0.1 mm}_{\beta}\gamma_{kn}
-\frac 12\Pi^s\hspace{0.1 mm}_nR^r\hspace{0.1 mm}_{\alpha, s}N^{\alpha \beta}R^k\hspace{0.1 mm}_{\beta}\gamma_{km}.
\label{vega}
\end{eqnarray}
The last term of (\ref{gamewarden}), using  the modified connection (\ref{vega}), becomes:
\begin{equation}
-\frac i2S[\phi]_{,j}\Delta [\phi ]^{nm} \bigg(\Gamma ^j\hspace{0.1 mm}_{qs}\Pi^q\hspace{0.1 mm}_m\Pi^s\hspace{0.1 mm}_n
-R^j\hspace{0.1 mm}_{\alpha, n} N^{\alpha \beta}R^k\hspace{0.1 mm}_{\beta}\gamma_{km}
-\Pi^{s}\hspace{0.1mm}_nR^j\hspace{0.1 mm}_{\alpha, s} N^{\alpha \beta}R^k\hspace{0.1 mm}_{\beta}\gamma_{km}
\bigg).
\label{vildtfoged}
\end{equation}
Picking a gauge where the propagator is restricted by the equation:
\begin{equation}
R^k\hspace{0.1 mm}_{\alpha}\gamma _{kn}\Delta [\phi ]^{nm}=0
\label{basilosaurus}
\end{equation}
(the Landau-De Witt gauge) this expression reduces to:
\begin{equation}
-\frac i2S[\phi]_{,j}\Delta [\phi ]^{nm} \Gamma ^j\hspace{0.1 mm}_{mn}.
\label{wildfang}
\end{equation}
This is the gauge used in \cite{Toms}, \cite{TomsII}, \cite{He}, \cite{Tang}, \cite{TomsIII}.
However, in order to prove gauge fixing independence of a gauge theory one has to keep the full expression (\ref{vildtfoged}). In the following sections we use (\ref{vildtfoged}) for one-loop quantum gravity and for the Maxwell-Einstein system to lowest and also next-lowest order (in Appendix B) in the gravitational coupling $\kappa$.

It is next verified that the gauge dependence of (\ref{gamewarden}) has been eliminated by addition of the last term. Gauge dependence occurs through  the gauge fixing function $\chi ^\alpha$ and also through the matrix elements $c_{\alpha \beta}$, but it is sufficient to consider variation of $\chi ^\alpha$, since the arbitrariness connected to  $c_{\alpha \beta}$ can be absorbed in the gauge fixing function.

From (\ref{gamewarden}) one finds:
\begin{equation}
\frac{\delta}
{\delta \chi^{\alpha}\hspace{0.1 mm}_{,j}}( -\frac i2 {\rm Tr} \log   \Delta[\phi])
= ic_{\alpha \beta}\chi^\beta\hspace{0.1 mm}_{,k}\Delta ^{kj}[\phi].
\label{sonnykarensdal}
\end{equation}
Assuming here and henceforth that the Ward identity (\ref{royalacademy}) can be applied  one gets from (\ref{sonnykarensdal}):
\begin{eqnarray}&&
\frac{\delta}{\delta \chi^{\alpha}\hspace{0.1 mm}_{,j}}( -\frac i2 {\rm Tr} \log   \Delta[\phi])
= i(Q^{-1})^{\delta}\hspace{0.1 mm}_\alpha(R^i\hspace{0.1 mm}_\delta+ S[\phi]_{,k}R^k\hspace{0.1 mm}_{\delta,l}\Delta[\phi]^{lj}).
\label{sonnydiamond}
\end{eqnarray}
where in (\ref{sonnydiamond}) the first term on the right-hand   is cancelled by the ghost term derivative:
\begin{equation}
\frac{\delta}{\delta \chi^{\alpha}\hspace{0.1 mm}_{,j}}(i{\rm Tr} \log Q^{-1})=- i(Q^{-1})^{\delta}\hspace{0.1 mm}_\alpha R^j\hspace{0.1 mm}_\delta.
\label{karensdal}
\end{equation}
(this could  be upset by the regularization scheme).
Also we find:
\begin{equation}
\frac {\delta}{\delta \chi ^\alpha \hspace{0.1 mm}_{,j}}\Delta [\phi]^{nm}
\simeq -\Delta[\phi]^{mj}(Q^{-1})^{\gamma}\hspace{0.1 mm}_\alpha R^n\hspace{0.1 mm}_\gamma
-\Delta[\phi_0]^{nj}(Q^{-1})^{\gamma}\hspace{0.1 mm}_\alpha R^m\hspace{0.1 mm}_\gamma\label{aladdin}
\end{equation}
by the Ward identity (\ref{royalacademy}) and where terms involving $S[\phi]_{,k}$ were disregarded, and thus we get from (\ref{vildtfoged}) after some manipulations and using (\ref{eliezer}) and (\ref{slotsherren}) and also the structure relations (\ref{derby}) and the gauge invariance of the classical action:
\begin{equation}
\frac {\delta}{\delta \chi^{\alpha}\hspace{0.1 mm}_{,j}}(-\frac i2S[\phi]_{,m}\tilde \Gamma ^m\hspace{0.1 mm}_{kl}[\phi_0]\Delta [\phi ]^{lk})
\simeq- iS[\phi]_{,k} R^k\hspace{0.1 mm}_{\beta, n}\Delta[\phi_0]^{nj}(Q^{-1})^{\beta}\hspace{0.1 mm}_\alpha.
\label{kahyasi}
\end{equation}
The sum of (\ref{sonnydiamond}), (\ref{karensdal}) and (\ref{kahyasi}) vanishes (the partial cancellation between (\ref{karensdal}) and (\ref{kahyasi})   could again be upset by the regularization scheme).

In summary, it was verified that the Vilkovisky construction as expected  formally removes the gauge parameter dependence of the effective action at one-loop order. In the course of this proof, conditions for the regularization scheme to be used  were obtained: It should  not upset the  cancellation between  (\ref{sonnydiamond}), (\ref{karensdal}) and (\ref{kahyasi}). It will be found in the following sections for the cases of quantum gravity and the Einstein-Maxwell system that these requirements are nontrivial and indeed are violated  by quartic and quadratic divergences in the Schwinger-DeWitt proper time representation of the effective action with a lower cut-off in the proper time variable.

\section{Pure quantum gravity in the one-loop approximation}

In this section we investigate the gauge parameter dependence of the one-loop effective action of pure quantum gravity with an arbitrary background metric, using the Schwinger-DeWitt proper time representation.   The one-loop effective action in a general field theory is determined from (\ref{gamewarden}), where in the proper time representation:
  \begin{equation}
  -\frac i2 {\rm Tr} \log \Delta[\phi]=  -\frac i2 {\rm Tr} \int _0^\infty \frac{d\tau}{\tau}e^{i\tau\Delta^{-1}[\phi]}
   \label{roxanne}
  \end{equation}
  where $\tau$ is the proper time, with a corresponding expression for the ghost contribution $i{\rm Tr}\log Q^{-1}$.
 This method is convenient because of the Campbell-Baker-Hausdorff identity:
\begin{equation}
\delta e^A=\int _0^1dt e^{tA}\delta Ae^{(1-t)A}
\label{CBH}
\end{equation}
with $A$ an arbitrary operator and $\delta A$ an infinitesimal variation that does not commute with $A$; this identity allows a perturbative expansion of the effective action.  (\ref{roxanne}) is an exact but possibly divergent representation of the effective action, where a regularization is achieved by a modification of the proper time integral at the lower end.

 The metric tensor $g_{\mu \nu}$ is split according to:
 \begin{equation}
  g_{\mu \nu}\rightarrow g_{\mu \nu}+\kappa h_{\mu \nu}
 \label{quant}
 \end{equation}
 with $g_{\mu \nu}$ a classical background metric field, while $h_{\mu \nu}$ is the quantum fluctuation field.  
 A coordinate transformation
  implies for  $h_{\mu\nu}$:
   \begin{equation}
 h_{\mu \nu}\rightarrow  h_{\mu \nu}+\xi_{\mu;\nu}+\xi_{\nu;\mu}+\kappa(h_{\mu\lambda}\xi^\lambda\hspace{0.1 mm}_{;\nu}+h_{\nu\lambda}\xi^\lambda \hspace{0.1 mm}_{;\mu}+\xi^\lambda h_{\mu\nu;\lambda})+O(\kappa^2).
 \label{supergauge}
 \end{equation}
 Here and elsewhere, covariant derivative is indicated by a semicolon, and if the covariant derivative is with respect to a variable $x'$, then the index following the semicolon carries a prime, etc..
 The Hilbert action (\ref{Hilbert}) has the linear term in  $h_{\mu \nu}$:
 \begin{equation}
 S^{(1)}_{H}=-\frac 1\kappa\int d^4x\sqrt{-g}h_{\mu \nu} {\cal G}^{\mu \nu} 
 \label{mousopenso}
 \end{equation}
  with $ {\cal G}^{\mu \nu}=R^{\mu \nu}-\frac 12 g^{\mu \nu}R$ the Einstein tensor of the background metric, where $R^{\mu\nu}$ the Ricci tensor and $R=g_{\mu \nu}R^{\mu \nu}$ the curvature scalar, and also the following quadratic term in $h_{\mu \nu}$:
 \begin{eqnarray}&&
S^{(2)}_{H}=\frac 12\int d^4x\sqrt{-g}\bigg(-\frac 12 h^\lambda \hspace{0.1 mm}_{\lambda }{\cal G}^{\mu \nu}h_{\mu \nu}
 +h^{\mu}\hspace{0.1 mm} _{ \rho}{\cal G}^{\rho \nu}h_{\mu \nu}+R^{\mu \lambda}h^\nu \hspace{0.1 mm}_{\lambda }h_{\mu \nu}-\frac 12 h^\mu \hspace{0.1 mm}_{\mu }R^{\lambda \rho}h_{\lambda \rho}
 \nonumber\\&&
 +\frac 12 h_{\mu \nu}g^{\mu \nu}(h^{\lambda \rho}\hspace{0.1 mm}_{;\lambda;\rho}-h^\lambda \hspace{0.1 mm}_{\lambda;}\hspace{0.1 mm}^{\rho}\hspace{0.1 mm}_{;\rho}) -\frac 12h^{\mu\nu} (h_{\mu \lambda;\nu;}\hspace{0.1 mm}^{\lambda}+h_{\nu \lambda;\mu;}\hspace{0.1 mm}^{\lambda}-h_{\mu \nu;}\hspace{0.1 mm}^{\lambda}\hspace{0.1 mm}_{;\lambda}-h^{\lambda}\hspace{0.1 mm}_{\lambda;\mu;\nu})\bigg).
 \label{Hamilton}
 \end{eqnarray}
In order to quantize the gravitational field one adds to (\ref{Hamilton}) a gauge breaking term:
 \begin{equation}
 S_{GB}=-\frac 12\frac 1\alpha\int d^4x\sqrt{-g}(h_{\mu \nu;}\hspace{0.1 mm}^\nu-\frac 12 g^{\nu \sigma}h_{\nu \sigma;\mu})g^{\mu \tau}(h_{\tau \lambda;}\hspace{0.1 mm}^\lambda-\frac 12g^{\lambda \rho}h_{\lambda \rho;\tau})
 \label{GB}
 \end{equation}
 where the gauge parameter $\alpha $ for simplicity is taken positive.
 The gauge breaking term (\ref{GB}) necessitates the Faddeev-Popov ghost action:
 \begin{equation}
 S_{FP}=\frac{1}{\sqrt{\alpha}}\int d^4x\sqrt{-g}\bar \xi ^\mu(\xi_{\mu;\nu;}\hspace{0.1 mm}^\nu+R_{\mu \nu}\xi^\nu)
 +O(\kappa ).
 \label{ghost}
 \end{equation}

 Here the factor $\frac{1}{\sqrt{\alpha}}$ in front, which usually is disregarded or removed by a rescaling of the ghost fields, is of crucial importance for the analysis of quadratic divergences. This factor   occurs in the ghost determinant and should hence also be taken along into the representation of this determinant that corresponds to (\ref{roxanne}). When these expressions are regularized by a lower cut-off in $\tau$ then a rescaling of the $\tau$-parameter by a factor $\sqrt{\alpha}$  in the expression for the ghost field determinant means that a different cut-off is used for the ghost determinant and the gauge field determinant.
 
 Also, leaving out the factor $\frac{1}{\sqrt \alpha}$ would give an artificial dependence of the path integral on the gauge-fixing parameter  as pointed out  in \cite{vanNieuw} in the case of Abelian  gauge fields (see especially the Ward identity argument in (2.18)-(2.19) of \cite{vanNieuw}). One could alternatively use an additional  ghost \cite{DeWittIII}, \cite{Kallosh} to remove this factor  from the ghost action, but the additional ghost would be non-propagating in this case, which  would make this procedure less convenient for actual calculations. 
 
The one-loop effective  action of quantum gravity is by (\ref{roxanne}), disregarding for a moment the ghost contribution:
    \begin{equation}
  \Gamma_{\rm gr} ^{[1]}=-\frac {i}{2}\int_0^\infty  \frac{d\tau}{\tau}\int d^4x(\frac 12 h^\alpha\hspace{0.1 mm}_{\mu \nu,}\hspace{0.1 mm}^{\mu \nu}(x,x;\tau)-\frac 14h^\alpha\hspace{0.1 mm}_\mu \hspace{0.1 mm}^\mu\hspace{0.1 mm}_{,\nu}\hspace{0.1 mm}^\nu(x,x;\tau))
  \label{cyrano}
  \end{equation}
  where the heat kernel \footnote{It is somewhat misleading to refer to this quantity as a heat kernel; this requires the proper time variable $\tau $ to be imaginary whereas  it is assumed  real here. However, we shall continue to use this name for simplicity.} $h^\alpha_{\mu \nu, \xi'\eta'}(x,x';\tau)$ is determined by the differential equation according to (\ref{Hamilton})-(\ref{GB}):
   \begin{eqnarray}&&
 i\frac{\partial}{\partial \tau}h^\alpha_{\mu \nu,\xi' \eta'}(x,x';\tau)+h^\alpha_{\mu \nu,\xi' \eta'}(x,x';\tau)_{;\sigma;}\hspace{0.1 mm}^\sigma-\frac 12X_{\mu\nu}\hspace{0.1 mm}^{\lambda \rho} h^\alpha_{\lambda \rho, \xi'\eta'}(x,x';\tau)
 \nonumber\\&&
-2R^{\lambda} \hspace{0.1 mm}  _ { \mu \nu} 
 \hspace{0.1 mm}^{ \rho}h^\alpha_{\lambda \rho, \xi'\eta'}(x,x';\tau) 
 \nonumber\\&&
 -(1-\frac 1\alpha)(h^\alpha_{\mu \lambda, \xi'\eta'}(x,x';\tau)_{;}\hspace{0.1 mm}^\lambda\hspace{0.1 mm}_{;\nu}+h^\alpha_{\nu \lambda, \xi'\eta'}(x,x';\tau)_{;}\hspace{0.1 mm}^\lambda\hspace{0.1 mm}_{;\mu}-
 g^{\lambda \rho}h^\alpha_{ \lambda\rho, \xi'\eta'}(x,x';\tau)_{;\mu;\nu})
 \nonumber\\&&
=0
 \label{exquisite}
 \end{eqnarray} 
 with the boundary condition:
  \begin{equation}
 h^\alpha_{\mu \nu, \xi'\eta'}(x,x';0)=(g_{\mu \xi'}g_{\nu \eta'}+g_{\nu \xi'}g_{\mu \eta'}-g_{\mu \nu}g_{\xi' \eta'})\delta (x,x')
 \label{burglary}
 \end{equation}
 and where:
 \begin{equation}
 \frac 12X_{\mu \nu }\hspace{0.1 mm}^{\lambda \rho}=R_{\mu \nu}g^{\lambda\rho}+g_{\mu \nu}{\cal G}^{\lambda \rho}-\delta _\mu \hspace{0.1 mm}^\lambda {\cal G}_\nu \hspace{0.1 mm}^\rho-\delta _\nu \hspace{0.1 mm}^\lambda {\cal G}_\mu \hspace{0.1 mm}^\rho+2R^{\lambda} \   _ { \mu \nu} \ ^{ \rho}.
 \label{uffehinspage}
 \end{equation}

  The ghost contribution to the effective action is by (\ref{ghost}):
 \begin{equation}
 \Gamma ^{[1]}_{\rm gh}=i\int _0^\infty \frac{d\tau}{\tau}\int d^4xh_{\rm gh, \mu }\hspace{0.1 mm}^\mu(x, x;\frac{1}{\sqrt \alpha}\tau)
 \label{roxolaner}
 \end{equation}
  with:
  \begin{equation}
 i\frac{\partial}{\partial \tau}h_{{\rm gh}, \mu , \xi'}(x,x';\tau)+h_{{\rm gh}, \mu , \xi'}(x,x';\tau)_{;\sigma;}\hspace{0.1 mm}^\sigma+R_\mu \hspace{0.1 mm}^\nu h_{{\rm gh}, \nu, \xi'}(x, x';\tau)=0
 \label{lindberg}
 \end{equation}
 and:
   \begin{equation}\hspace{0.1 mm}
h_{{\rm gh}, \mu , \xi'}(x,x';0)=g_{\mu \xi'}\delta (x,x').
 \label{anna}
 \end{equation}
 This heat kernel fulfills the following important relation:
  \begin{equation}
 h_{{\rm gh},\mu \xi'} (x,x';\tau_1+\tau_2)=\int d^4x''h_{{\rm gh}, \mu }\hspace{0.1 mm}^{\sigma''}(x,x'';\tau_1)h_{{\rm gh}, \sigma'' \xi'}(x'',x'; \tau_2)
 \label{semigroup}
 \end{equation}
by (\ref{ghost}).  An analogous relation holds for the graviton heat kernel.

One should notice the square root of the gauge parameter $\alpha$ in (\ref{roxolaner}). Formally this dependence on $\alpha$ can, as mentioned above, be removed by a rescaling of the integration variable $\tau$. When a lower cut-off is introduced in the integral, however, the quartic and quadratic divergences will depend on $\alpha$.

 The dependence of the heat kernel  $h^\alpha_{\mu \nu, \xi'\eta'}(x,x';\tau)$ on the gauge parameter $\alpha$  is next determined. First it is shown that the heat kernel obeys a Ward identity. Heat kernel Ward identities were considered previously in \cite{vanNieuw}, \cite{NKN}.
From (\ref{exquisite}) follows:
  \begin{eqnarray}&&
  i\frac{\partial}{\partial \tau}(h^\alpha _{\mu \nu, \xi'\eta'}(x,x';\tau)_{;}\hspace{0.1 mm}^\mu-\frac 12h^\alpha_{\mu }\hspace{0.1 mm}^{\mu} \hspace{0.1 mm} _ {, \xi'\eta'}(x,x';\tau)_{;}\hspace{0.1 mm}^\nu)
  \nonumber\\&&
  +\frac 1\alpha(h^\alpha_{\mu \nu, \xi'\eta'}(x,x';\tau)_{;}\hspace{0.1 mm}^\mu-\frac 12h^\alpha_{\mu }\hspace{0.1 mm}^{\mu} \hspace{0.1 mm} _ {, \xi'\eta'}(x,x';\tau)_{;}\hspace{0.1 mm}^\nu)\hspace{0.1 mm}_{;\sigma;}\hspace{0.1 mm}^\sigma
  \nonumber\\&&
  +\frac 1\alpha R_\nu \hspace{0.1 mm}^\lambda (h^\alpha_{\mu \lambda, \xi'\eta'}(x,x';\tau)_{;}\hspace{0.1 mm}^\mu-\frac 12h^\alpha_{\mu }\hspace{0.1 mm}^{\mu} \hspace{0.1 mm} _ {, \xi'\eta'}(x,x';\tau)_{;}\hspace{0.1 mm}^\lambda)
  \nonumber\\&&
 =-{\cal G}^{\lambda \rho}(2h^\alpha_{\lambda \nu ,\xi'\eta'}(x,x';\tau)_{;\rho}-h^\alpha_{\lambda \rho, \xi'\eta'}(x,x';\tau)\hspace{0.1 mm}_{;\nu})
 \label{Dettmers}
 \end{eqnarray}
 with the solution:
 \begin{eqnarray}&&
  h^\alpha _{\mu \nu,\xi' \eta'}(x,x';\tau)_{;}\hspace{0.1 mm}^\mu -\frac 12h^\alpha _{\mu } \hspace{0.1 mm}  ^{\mu}  \hspace{0.1 mm}  _ {,\xi' \eta'}(x,x';\tau)_{;\nu }
  \nonumber\\&&
  =-h_{{\rm gh},\nu,\eta'}(x,x';\frac 1\alpha\tau)_{;\xi'}-h_{{\rm gh}, \nu,\xi'}(x,x';\frac 1\alpha\tau)_{;\eta'}
  \nonumber\\&&
  +i\tau \int d^4x''\int _0^1dth_{{\rm gh}, \nu }\hspace{0.1 mm}^{\sigma''}(x.x'';\frac 1\alpha t\tau){\cal G}^{\omega'' \delta''}(x'')\bigg(2h^\alpha _{\omega ''\sigma'' , \xi'\eta'}(x'',x';(1-t)\tau)\hspace{0.1 mm}_{;\delta''}
  \nonumber\\&&
  -h^\alpha _{\omega''\delta''}\hspace{0.1 mm}, _{\xi'\eta'}(x'',x';(1-t)\tau)\hspace{0.1 mm}_{;\sigma''}\bigg).
 \label{kalmar}
 \end{eqnarray}

 From (\ref{exquisite}) also follows a differential equation for the function  $\alpha \frac{\partial}{\partial \alpha}  h^\alpha _{\mu \nu,\xi' \eta'}(x,x';\tau)$:
   \begin{eqnarray}&&
 i\frac{\partial}{\partial \tau}(\alpha \frac{\partial}{\partial \alpha} h^\alpha_{\mu \nu,\xi' \eta'}(x,x';\tau))+\alpha \frac{\partial}{\partial \alpha} h^\alpha_{\mu \nu,\xi' \eta'}(x,x';\tau)_{;\sigma;}\hspace{0.1 mm}^\sigma-\frac 12X_{\mu\nu}\hspace{0.1 mm}^{\lambda \rho} \alpha \frac{\partial}{\partial \alpha} h^\alpha_{\lambda \rho, \xi'\eta'}(x,x';\tau)
\nonumber\\&&
 -(1-\frac 1\alpha)(\alpha \frac{\partial}{\partial \alpha} h^\alpha_{\mu \lambda, \xi'\eta'}(x,x';\tau)_{;}\hspace{0.1 mm}^\lambda\hspace{0.1 mm}_{;\nu}+\alpha \frac{\partial}{\partial \alpha} h^\alpha_{\nu \lambda, \xi'\eta'}(x,x';\tau)_{;}\hspace{0.1 mm}^\lambda\hspace{0.1 mm}_{;\mu}-
 g^{\lambda \rho}\alpha \frac{\partial}{\partial \alpha} h^\alpha_{ \lambda\rho, \xi'\eta'}(x,x';\tau)_{;\mu;\nu})
 \nonumber\\&&
=\frac 1\alpha(h^\alpha_{\mu \lambda, \xi'\eta'}(x,x';\tau)_{;}\hspace{0.1 mm}^\lambda\hspace{0.1 mm}_{;\nu}+h^\alpha_{\nu \lambda, \xi'\eta'}(x,x';\tau)_{;}\hspace{0.1 mm}^\lambda\hspace{0.1 mm}_{;\mu}-
 g^{\lambda \rho}h^\alpha_{ \lambda\rho, \xi'\eta'}(x,x';\tau)_{;\mu;\nu}).
 \label{elegant}
 \end{eqnarray} 
 This equation is solved in the same way as (\ref{Dettmers}); the result is:
 \begin{eqnarray}&&
 \alpha \frac{\partial}{\partial \alpha}  h^\alpha _{\mu \nu,\xi' \eta'}(x,x';\tau)
 \nonumber\\&&
 =\frac 1\alpha i\tau \int d^4x''\int_0^1dt(h^\alpha _{\mu \nu, \lambda ''\rho''}(x, x''; t\tau)_;\hspace{0.1 mm}^{\lambda ''}
 -\frac 12h^\alpha _{\mu \nu, \lambda ''}\hspace{0.1 mm}^{\lambda''}(x, x''; t\tau)_{;\rho''})g^{\rho''\omega''}(x'')
 \nonumber\\&&
 (h^\alpha _{\sigma ''\omega'', \xi' \eta'}(x'', x'; (1-t)\tau)_;\hspace{0.1 mm}^{\sigma ''}
 -\frac 12h^\alpha _{\sigma ''}\hspace{0.1 mm}^{\sigma''}\hspace{-1 mm}_{, \xi' \eta'}(x'', x'; (1-t)\tau)_{;\omega''}).
 \label{ricciardelli}
 \end{eqnarray}
 Using here (\ref{kalmar}) and disregarding the second term on the right-hand side one gets approximately:
 \begin{eqnarray}&&
 \alpha \frac{\partial}{\partial \alpha}  h^\alpha _{\mu \nu,\xi' \eta'}(x,x';\tau)
 \simeq -\frac 1\alpha i\tau \int d^4x''\int _0^1 dt(h_{{\rm gh}\mu \rho''}(x, x'';t\frac 1\alpha \tau)_{;\nu}+h_{{\rm gh}\nu \rho''}(x, x'';t\frac 1\alpha \tau)_{;\mu})g^{\rho''\omega''}(x'')
  \nonumber\\&&
   (h^\alpha _{\sigma ''\omega'', \xi' \eta'}(x'', x'; (1-t)\tau)_;\hspace{0.1 mm}^{\sigma ''}
 -\frac 12h^\alpha _{\sigma ''}\hspace{0.1 mm}^{\sigma''}\hspace{-1 mm}_{, \xi' \eta'}(x'', x'; (1-t)\tau)_{;\omega''}).
  \label{tinaogmarina}
 \end{eqnarray}
  Using again (\ref{kalmar}) and also (\ref{semigroup}) one gets a further approximation:
 \begin{eqnarray}&&
 \alpha \frac{\partial}{\partial \alpha}  h^\alpha _{\mu \nu,\xi' \eta'}(x,x';\tau)
 \simeq  \frac 1\alpha i\tau ((h_{{\rm gh}, \nu,\eta'}(x,x';\frac 1\alpha\tau))_{;\mu;\xi'}+(h_{{\rm gh}, \nu,\xi'}(x,x';\frac 1\alpha\tau))_{;\mu;\eta'}
   \nonumber\\&&
   +(h_{{\rm gh}, \mu,\eta'}(x,x';\frac 1\alpha\tau))_{;\nu;\xi'}+(h_{{\rm gh}, \mu,\xi'}(x,x';\frac 1\alpha\tau))_{;\nu;\eta'}).
 \label{lammelaar}
 \end{eqnarray}

  Combining (\ref{ricciardelli}) with (\ref{cyrano}) and   the relation analogous to  (\ref{semigroup}) for the graviton heat kernel  the dependence of the effective action of the gauge parameter $\alpha$ can be found:
 \begin{eqnarray}&&
 \alpha \frac{\partial}{\partial \alpha} \Gamma_{\rm gr}^{[1]}=\frac 12 \frac 1\alpha \int_0^\infty d\tau \int d^4 xg^{\nu \eta'}(h^\alpha _{\mu \nu, \xi'\eta'}(x, x';\tau)_{;}^\mu \hspace{0.1 mm}_{;}\hspace{0.1 mm}^{\xi'}
 -\frac 12h^\alpha _{\mu}\hspace{0.1 mm}^{\mu}\hspace{0.1 mm}_{,\xi'\eta'}(x, x';\tau)_{;\nu;}\hspace{0.1 mm}^{\xi'}  
 \nonumber\\&&
-\frac 12h^\alpha _{\mu \nu, \xi'}\hspace{0.1 mm}^{\xi'}(x, x';\tau)_{;}^\mu \hspace{0.1 mm}_{;\eta'} +\frac 14 h^\alpha _{\mu}\hspace{0.1 mm}^\mu \hspace{0.1 mm}_{  \xi'}\hspace{0.1 mm}^{\xi'}(x, x';\tau)_{;\nu ;\eta'}).
 \label{kulmule}
 \end{eqnarray}
 By means of  (\ref{kalmar}) one gets from (\ref{kulmule}) two terms that correspond precisely to the two terms of (\ref{sonnydiamond}): 
 \begin{eqnarray}&&
  \alpha \frac{\partial }{\partial \alpha}\Gamma_{{\rm gr}, I} ^{[1]}= \frac 12 \frac 1\alpha \int_0^\infty d\tau\int d^4x g^{\mu \eta'}(-h_{{\rm gh}, \mu,\eta'}(x,x';\frac 1\alpha\tau)_{;\xi' ;}\hspace{0.1 mm}^{\xi'}-h_{{\rm gh}, \mu,\xi'}(x,x';\frac 1\alpha\tau)_{;\eta ' ;}\hspace{0.1 mm}^{\xi'}
   \nonumber\\&&
+h_{{\rm gh}, \mu,\xi'}(x,x';\frac 1\alpha\tau)_{;}\hspace{0.1 mm}^{\xi'}\hspace{0.1 mm}_{ ;\eta'})\mid_{x'\rightarrow x}
\nonumber\\&&
=\frac i2 \int d^4xg^{\mu \eta'}\int _0^\infty d\tau\frac{\partial}{\partial \tau}h_{{\rm gh}, \mu,\eta'}(x,x;\frac 1\alpha\tau)
  \label{bergerac}
  \end{eqnarray}
 where (\ref{lindberg}) was used in the last step,  and:
  \begin{eqnarray}&&
 \alpha \frac{\partial}{\partial \alpha}\Gamma_{{\rm gr}, II} ^{[1]}=  i\frac 1\alpha\int d^4x\int d^4x' \int_0^\infty  \tau d\tau\int _0^1 dth_{{\rm gh},}\hspace{0.1 mm}^{\mu \sigma'}(x, x';t\frac 1\alpha \tau)
   {\cal G}^{\omega '\delta'}(x')
   \nonumber\\&&
  ( h^\alpha_{\omega'\sigma',\mu \nu}(x', x;(1-t)\tau)_{;\delta'}\hspace{0.1 mm}^{;\nu}-\frac 12  h^\alpha_{\omega'\sigma',\nu}\hspace{0.1 mm}^{ \nu}(x', x;(1-t)\tau)_{;\delta'}\hspace{0.1 mm}_{;\mu}
   \nonumber\\&&
   -\frac 12(h^\alpha_{\omega'\delta',\mu \nu}(x', x;(1-t)\tau)_{;\sigma'}\hspace{0.1 mm}^{;\nu}
  -\frac 12  h^\alpha_{\omega'\delta',\nu}\hspace{0.1 mm}^{ \nu}(x', x;(1-t)\tau)_{;\sigma'}\hspace{0.1 mm}_{;\mu})).
\label{lamborghini}
\end{eqnarray}

 (\ref{bergerac}) is an integral of a total derivative. It is considered in connection with the ghost contribution to the effective action (\ref{roxolaner}) from which one finds:
  \begin{equation}
  \alpha \frac{\partial}{\partial \alpha} \Gamma^{[1]}_{\rm gh}=-\frac i2\int d^4x \int _0^\infty d\tau \frac{\partial}{\partial \tau}h_{\rm gh, \mu }\hspace{0.1 mm}^\mu (x, x;\frac{1}{\sqrt \alpha}\tau)
  \label{shehrezade}
  \end{equation}
  that has the same form as (\ref{bergerac}), with a sign change and a different dependence on $\alpha$. It would cancel with (\ref{bergerac}) by a rescaling of the the proper time variable $\tau$. 
  Indeed, it is an example of the partial cancellation between the general expressions (\ref{sonnydiamond}) and (\ref{karensdal}). However, a careful examination of the quartic and quadratic divergences is necessary here. Using (\ref{malacca}) and (\ref{shorthorn}) one gets from (\ref{bergerac}):
  \begin{equation}
    \alpha \frac{\partial }{\partial \alpha}\Gamma _{{\rm gr}, I}^{[1]}\simeq \frac 12 \frac{1}{16\pi^2}\int d^4x\sqrt{-g}(-\frac{4\alpha^2}{\tau^2}+\frac 53R\frac{i\alpha}{\tau})\mid _{\tau\simeq 0}
  \label{demongeot}
  \end{equation}
  where a conventional cut-off $\Lambda$ can be introduced by the substitution 
  \begin{equation}
  \tau \simeq -i\frac{1}{\Lambda ^2}.
  \label{gulliver}
  \end{equation}
  From (\ref{shehrezade}) one gets:
  \begin{equation}
   \alpha \frac{\partial}{\partial \alpha} \Gamma^{[1]}_{\rm gh}\simeq -\frac 12 \frac{1}{16\pi^2}\int d^4x\sqrt{-g}(-\frac{4\alpha}{\tau^2}+\frac 53R\frac{i\sqrt \alpha}{\tau})\mid _{\tau\simeq 0}
  \label{mylene}
  \end{equation}
  that does not cancel with (\ref{demongeot}) at general $\alpha$. The two expressions contain both quadratic and quartic divergences. \footnote{Quartic divergences have the same order of magnitude as the divergences related to the integral measure in the Lagrangian version of the path integral and should be considered in connection with these. I am grateful to a referee for stressing this point.}

  The  expression (\ref{lamborghini}) is also quadratically divergent, and  its divergent part is determined by first using the Ward identity (\ref{kalmar}), disregarding the second term on the right hand side, and also the relation (\ref{semigroup}):
   \begin{equation}
   \alpha \frac{\partial}{\partial \alpha}\Gamma_{{\rm gr}, II} ^{[1]}\simeq -i\frac 1\alpha \int d^4x\int_0^\infty \tau d\tau {\cal G}^{\omega '\delta'}(x)h_{{\rm gh}, \sigma'\mu}(x',x;  \frac 1\alpha \tau)_{;\omega';\delta'}\mid _{x'\rightarrow x}g^{\mu \sigma'}.
  \label{greengables}
  \end{equation}
  By means of (\ref{malaya}) one then obtains from (\ref{greengables}):
  \begin{equation}
   \alpha \frac{\partial}{\partial \alpha}\Gamma_{{\rm gr}, II} ^{[1]}\simeq - 2\alpha^2\frac{i}{16\pi^2}\frac{1}{\tau}\mid _{\tau\simeq 0}\int d^4x\sqrt{-g}R.
  \label{kilkenney}
  \end{equation}

  The gauge dependence of the one-loop effective action of pure gravity (\ref{cyrano}) is contained in the  expressions (\ref{bergerac}), which is formally cancelled by the ghost contribution (\ref{shehrezade}), as well as in  (\ref{lamborghini})  that vanishes in an Einstein-flat space-time with ${\cal G}^{\mu \nu}=0$ and where the gauge parameter dependence is expected to be formally removed by the Vilkovisky construction. In the presence of matter fields we expect that the gauge dependent part of the effective action still contains (\ref{lamborghini}), with the replacement ${\cal G} ^{\mu \nu}\rightarrow  {\cal G} ^{\mu \nu}-{\cal T} ^{\mu \nu}$, where  ${\cal T}^{\mu \nu}$ is the energy-momentum tensor of the background matter field.

  We then work out the details of the Vilkovisky construction in
 quantum gravity. This topic  has previously been considered in \cite{Barvinsky}, \cite{Kunstatter}, \cite{Odintsov}.
 We here use the proper-time representation of the effective action,  such that  the formal cancellation of  (\ref{kahyasi}) with (\ref{sonnydiamond}) and (\ref{karensdal})  can be investigated on the regularized level in this case.

  When the effective action is extended by Vilkovisky and De Witt's method to field configurations where the background field equations are not valid, new  terms are introduced in quantum gravity by  (\ref{vildtfoged}), with:
\begin{equation}
S_{, h_{\mu \nu}}=-\frac {1}{\kappa}{\cal G}^{\mu\nu}.
\label{mikkelmax}
\end{equation}
 In quantum gravity the field metric can be chosen as:
\begin{equation}
G_{h_{\mu \nu}(x), h_{\lambda' \rho'}(x')}=\frac 14\sqrt{-g}(g^{\mu \lambda'}g^{\nu \rho'}+g^{\nu \lambda'}g^{\mu \rho'}-g^{\mu \nu}g^{\lambda' \rho'})(x)\delta(x, x').
\label{bjarke}
\end{equation}
For gravitational fluctuations $h_{\mu\nu}$ the transformation (\ref{supergauge}) determines by means of  (\ref{eliezer}) and (\ref{lindberg}):
 \begin{equation}
 N^{\alpha \beta}\rightarrow N^{\xi_\mu(x)\xi _{\nu'} (x')}= i\alpha ^k\int _0^\infty d\tau \frac{1}{^4\sqrt{-g}}h_{{\rm gh}, \mu \nu'}(x, x';\alpha ^k\tau)\frac{1}{^4\sqrt{-g'}}
  \label{lloydbridges}
 \end{equation}
 with $k$ so far unspecified.

Using also (\ref{supergauge}) one now  finds:
\begin{eqnarray}&&
\frac i2S[\phi_0]_{,j}\Delta[\phi_0]^{nm}R^j\hspace{0.1 mm}_{\alpha, n} N^{\alpha \beta}R^k\hspace{0.1 mm}_{\beta }\gamma _{km}
\nonumber\\&&
\rightarrow 
\frac 12  \int d^4x\int d^4y \int d^4 z\int d^4 w\int d^4 u\int d^4 tS_{, h_{\mu \nu}(x)}R^{h_{\mu \nu}(x)}\hspace{0.1 mm}_{\xi_\omega(y), h_{\lambda \rho}(z)}
 \nonumber\\&&
 N^{\xi_\omega(y)\xi_\sigma (w)}
  R^{h_{\xi \eta}(u)}\hspace{0.1 mm}_{\xi_\sigma(w)}G_{\xi \eta, \alpha \beta}(u,t)
 <h_{\alpha \beta}(t)h_{\lambda \rho}(z)>
 \nonumber\\&&
\simeq - i \alpha^k\int d^4x\int d^4x' \int_0^\infty  \tau d\tau\int _0^1 dth_{{\rm gh},}\hspace{0.1 mm}^{\mu \sigma'}(x, x';t\alpha ^k \tau)
   {\cal G}^{\omega '\delta'}(x')
   \nonumber\\&&
  ( h^\alpha_{\omega'\sigma',\mu \nu}(x', x;(1-t)\tau)_{;\delta'}\hspace{0.1 mm}^{;\nu}-\frac 12  h^\alpha_{\omega'\sigma',\nu}\hspace{0.1 mm}^{ \nu}(x', x;(1-t)\tau)_{;\delta'}\hspace{0.1 mm}_{;\mu}
   \nonumber\\&&
   -\frac 12(h^\alpha_{\omega'\delta',\mu \nu}(x', x;(1-t)\tau)_{;\sigma'}\hspace{0.1 mm}^{;\nu}
  -\frac 12  h^\alpha_{\omega'\delta',\nu}\hspace{0.1 mm}^{ \nu}(x', x;(1-t)\tau)_{;\sigma'}\hspace{0.1 mm}_{;\mu})) 
\label{starman}
\end{eqnarray}
where the graviton propagator is: 
   \begin{equation}
 <h_{\mu \nu}(x)h_{\xi'\eta'}(x')>=\int_0^{\infty} d\tau \frac{1}{^4 \sqrt{-g}}h^\alpha_{\mu \nu, \xi'\eta'}(x,x';\tau)\frac{1}{ ^4 \sqrt{-g'}}.
 \label{Magenta}
 \end{equation}
 (\ref{starman})  vanishes in the Landau-DeWitt gauge (the limit $\alpha= 0$) by  the Ward identity (\ref{kalmar}), where only the first hand on the right-hand side is kept.
Requiring that (\ref{starman}) cancels  with  (\ref{lamborghini}) fixes $k$ at $-1$;  at other values of $k$ there is formally  still cancellation as seen by introducing $\tau_1=t\tau, \tau_2=(1-t)\tau$ and performing a rescaling of $\tau_1$. However, this argument is upset by the quadratic divergences of the two expressions.
Evaluating the quadratic divergence of (\ref{starman}) at general $k$ in the same way as (\ref{kilkenney}) by means of (\ref{semigroup}) one finds:
   \begin{equation}
   -\frac{1}{16\pi^2}i\alpha^2\frac{1-\alpha^{-2(k+1)}}{1-\alpha ^{k+1}}\frac{1}{\tau}\mid _{\tau\simeq 0}\int d^4x\sqrt{-g}R
  \label{killarney}
  \end{equation}
 that only cancels (\ref{kilkenney}) at $k=-1$, showing that the requirement that (\ref{starman}) cancels with (\ref{lamborghini}) also for quadratic divergences is a nontrivial one.

The Christoffel connection in field space is:
\begin{eqnarray}&&
\Gamma ^{h_{\sigma \omega}(x)}\hspace{0.1 mm}_{h_{\mu \nu}(y), h_{ \lambda \rho}(z)}
\nonumber\\&&
=\frac 14\bigg(\delta _{(\sigma \omega)}\hspace{0.1 mm}^{(\mu\nu )}g^{\lambda \rho}+\delta _{(\sigma \omega)}\hspace{0.1 mm}^{(\lambda \rho )}g^{\mu \nu}
-\delta _{(\sigma \omega)}\hspace{0.1 mm}^{(\nu \rho)}g^{\mu\lambda }-\delta _{(\sigma \omega)}\hspace{0.1 mm}^{(\nu\lambda  )}g^{\mu \rho}-\delta _{(\sigma \omega)}\hspace{0.1 mm}^{(\mu \rho)}g^{\nu\lambda }
\nonumber\\&&
-\delta _{(\sigma \omega)}\hspace{0.1 mm}^{(\mu\lambda  )}g^{\nu \rho}
+ g^{(\mu \nu)(\lambda \rho)}g_{\sigma \omega}-\frac 12 g^{\mu \nu}g^{\lambda \rho}g_{\sigma \omega}\bigg)(y)\delta (x,y)\delta (y,z)
\label{esketjustrup}
\end{eqnarray}
with:
\begin{equation}
\delta _{(\mu\nu)}\hspace{0.1 mm}^{(\sigma \omega)}=\frac 12(\delta _\mu \hspace{0.1 mm}^\sigma \delta _\nu\hspace{0.1 mm}^\omega+\delta _\nu \hspace{0.1 mm}^\sigma \delta _\mu\hspace{0.1 mm}^\omega), \ g^{(\mu\nu)(\lambda \rho)}=\frac 12 (g^{\mu \lambda}g^{\nu\rho}+g^{\mu \rho}g^{\nu \lambda}).
\label{holmegaard}
\end{equation}
From (\ref{lloydbridges}) follows that the projection operator $\Pi ^m\hspace{0.1 mm}_n$ defined in (\ref{stamherren}) in this case is:
  \begin{eqnarray}&&
 \Pi^m\hspace{0.1 mm}_n\rightarrow \Pi^{h_{\mu \nu}(x)}\hspace{0.1 mm}_{h_{\lambda \rho}(y)}
 = \delta _{(\mu \nu)}\hspace{0.1 mm}^{(\lambda \rho)}\delta (x, y)
  \nonumber\\&&
  -i^4\sqrt{-g}\frac{1}{\sqrt \alpha}<(\xi_{\mu;\nu}+\xi_{\nu;\mu})(x)\frac 12 (\bar \xi^ \lambda\hspace{0.1 mm}_{;}\hspace{0.1 mm}^\rho +\bar \xi^ \rho\hspace{0.1 mm}_{;}\hspace{0.1 mm}^\lambda-g^{\lambda \rho}\bar \xi_{\sigma;}\hspace{0.1 mm}^\sigma)(y)>^4\sqrt{-g}
 \nonumber\\&&
 \label{carolina}
 \end{eqnarray}
 where the ghost propagator is:
  \begin{equation}
 <\xi_\mu(x)\bar \xi_\nu(y)>=\frac{1}{\sqrt \alpha} \int_0^{\infty} d\tau \frac{1}{^4 \sqrt{-g}}h_{{\rm gh}, \mu , \xi'}(x,x';\frac 1\alpha \tau)\frac{1}{ ^4 \sqrt{-g'}}.
 \label{askeladden}
 \end{equation}
 Applying this projection operator upon the graviton propagator one obtains the graviton propagator in the Landau-DeWitt gauge. From (\ref{vildtfoged}) one gets by (\ref{esketjustrup}):
\begin{eqnarray}&&
-\frac i2S[\phi_0]_{,m}\Gamma ^m\hspace{0.1 mm}_{kl}[\phi_0] \Pi^k\hspace{0.1 mm}_r\Delta [\phi_0 ]^{rs} \Pi ^l\hspace{0.1 mm}_s
\nonumber\\&&
\rightarrow \frac{1}{8}\int d^4x\sqrt{-g}{\cal G}^{\sigma\omega}(x)<h_{\mu \nu}(x)h_{\lambda \rho}(x)>
\nonumber\\&&
\bigg(\delta _{(\sigma \omega)}\hspace{0.1 mm}^{(\mu\nu )}g^{\lambda \rho}+\delta _{(\sigma \omega)}\hspace{0.1 mm}^{(\lambda \rho )}g^{\mu \nu}
-\delta _{(\sigma \omega)}\hspace{0.1 mm}^{(\nu \rho)}g^{\mu\lambda }-\delta _{(\sigma \omega)}\hspace{0.1 mm}^{(\nu\lambda  )}g^{\mu \rho}-\delta _{(\sigma \omega)}\hspace{0.1 mm}^{(\mu \rho)}g^{\nu\lambda }-\delta _{(\sigma \omega)}\hspace{0.1 mm}^{(\mu\lambda  )}g^{\nu \rho})
\nonumber\\&&
+ g^{(\mu \nu)(\lambda \rho)}g_{\sigma \omega}-\frac 12 g^{\mu \nu}g^{\lambda \rho}g_{\sigma \omega}\bigg)
\label{prokofiev}
\end{eqnarray}
with the  graviton propagator in the Landau-DeWitt gauge. This expression has no quadratic divergence in four dimensions by (\ref{cordova}).
There is no contribution to the effective action in this case from the final term in (\ref{vildtfoged}).

\section{The Maxwell-Einstein system}

  \subsection{Maxwell field in a curved background}
  
   The Maxwell field $A_\mu$  has the action:
 \begin{equation}
 S_M=\int d^4x\sqrt{-g}g^{\mu \lambda}g^{\nu \rho}(-\frac 14 F_{\mu \nu}F_{\lambda \rho}); \ F_{\mu \nu }=\partial _\mu A_\nu-\partial _\nu A_\mu.
 \label{Maxwell}
 \end{equation}
  Here $A_\mu$ can be considered a covariant vector with the following  transformation rule under infinitesimal coordinate transformations:
  \begin{equation}
  \delta  A_\mu =\kappa(\xi^\lambda\hspace{0.1 mm}_{;\mu}A_{\lambda}+\xi^\lambda A_{\mu; \lambda})+O(\kappa^2).
  \label{obligation}
  \end{equation} 
 (\ref{Maxwell}) gets  by the splitting (\ref{quant})    the additional term:
 \begin{equation}
  S_{\rm M}^{(1)}=\frac 1\kappa\int d^4x\sqrt{-g}h_{\mu \nu}T^{\mu \nu}
 \label{anaximander}
 \end{equation}
 where the energy-momentum tensor $T^{\mu \nu}$ is:
 \begin{eqnarray}&&
 T^{\mu \nu}=\frac {\kappa^2}{2}(F^{\mu \lambda}F^\nu \hspace{0.1 mm}_\lambda-\frac 14 g^{\mu \nu}F^{\lambda \rho}F_{\lambda \rho}).
 \label{anaximenes}
 \end{eqnarray}
   At second order in $\kappa$ one gets from (\ref{Maxwell}):
 \begin{eqnarray}&&
  S_{\rm M}^{(2)}=\kappa^2\int d^4x\sqrt{-g}h_{\omega \tau}h_{\iota \sigma}(\frac{1}{8}g^{\omega \tau}( F^\iota \hspace{0.1  mm}_\lambda F^{\sigma \lambda}-\frac 14g^{\iota \sigma} F^{\mu \nu} F_{\mu \nu})+\frac{1}{16}g^{\omega \iota}g^{\sigma \tau} F^{\mu \nu} F_{\mu \nu}
  \nonumber\\&&
 +\frac 18g^{\omega \tau} F^\iota \hspace{0.1  mm}_\lambda F^{\sigma \lambda}
 -\frac 14 F^{\omega \iota} F^{\tau \sigma}-\frac 12 g^{\sigma \omega} F^\iota \hspace{0.1  mm}_\lambda F^{\tau \lambda}).
 \label{mattgroening}
 \end{eqnarray}
 No splitting of the Maxwell field $A_\mu$ into a background field and an interacting field has been carried out yet.
  The gauge breaking action of the Maxwell field is:
 \begin{equation}
 S_{\rm{M, GB}}=\int d^4x\sqrt{-g}(-\frac 12\frac 1\beta(A_{\mu;} \hspace{0.1 mm}^\mu)^2)
 \label{Maxbreak}
 \end{equation}
 with the gauge parameter $\beta > 0$, and the corresponding ghost action:
 \begin{equation}
 S_{\rm{M, FP}}=\frac{1}{\sqrt{\beta}}\int d^4x\sqrt{-g}\bar c(c,_\mu+\kappa(\xi^\lambda\hspace{0.1 mm}_{;\mu}A_{\lambda}+\xi^\lambda A_{\mu; \lambda})) \hspace{0.1 mm}_{;}\hspace{0.1 mm}^\mu
 \label{Maxghost}
 \end{equation}
 with $c$ a scalar ghost and $\bar c$ the corresponding antighost.
   
 The photon heat kernel  $h^\beta_{ \mu, \xi'}(x, x';\tau)$ is defined by:
   \begin{eqnarray}&&
 i\frac{\partial}{\partial \tau}h^\beta_{\mu , \xi'}(x,x';\tau)+h^\beta_{ \mu , \xi'}(x,x';\tau)_{;\sigma;}\hspace{0.1 mm}^\sigma
 \nonumber\\&&
 -R_\mu \hspace{0.1 mm}^\nu h^\beta _{\nu, \xi'}(x, x';\tau)-(1-\frac 1\beta) h^\beta_{\nu,\xi'}(x,x';\tau)_{;}\hspace{0.1 mm}^\nu\hspace{0.1 mm}_{;\mu}=0
 \label{pontus}
 \end{eqnarray}
 where the boundary condition is:
   \begin{equation}\hspace{0.1 mm}
h^\beta_{ \mu , \xi'}(x,x';0)=g_{\mu \xi'}\delta (x,x').
 \label{frits}
 \end{equation}
Also the scalar heat kernel  $h(x, x';\tau)$ is defined by:
  \begin{equation}
 i\frac{\partial}{\partial \tau}h(x,x';\tau)+h(x,x';\tau)_{;\sigma;}\hspace{0.1 mm}^\sigma=0; \hspace{0.1 mm}
h(x,x';0)=\delta (x,x').
 \label{lufthanna}
 \end{equation}
 From (\ref{pontus}) follows:
  \begin{equation}
  i\frac{\partial}{\partial \tau}h^\beta_{ \mu , \xi'}(x,x';\tau)_{;}\hspace{0.1 mm}^\mu+\frac 1\beta h^\beta _{\mu , \xi'}(x,x';\tau)_{;}\hspace{0.1 mm}^\mu\hspace{0.1 mm}_{;\sigma;}\hspace{0.1 mm}^\sigma=0
  \label{myllymaeki}
  \end{equation}
 the solution of which is the following Ward identity:
  \begin{equation}
  h^\beta_{ \mu , \xi'}(x,x';\tau)_{;}^\mu=-h(x, x';\frac 1\beta \tau)_{;\xi'}
  \label{monpetit}
  \end{equation}
  obtained by comparison of  (\ref{lufthanna}) and (\ref{myllymaeki}) and by taking the boundary conditions into account. Also (\ref{monpetit}) combined with (\ref{lufthanna}) imply (cp.  \cite{Endo} eq. (2.23)):
\begin{equation}
h^\beta_{\mu,\xi'}(x,x';\tau)=h_{\mu,\xi'}(x,x';\tau)-i\int _\tau^{\frac 1\beta \tau}d\tau'h(x,x';\tau')_{;\mu;\xi'}
\label{Kermes}
\end{equation}
where $h_{\mu,\xi'}(x,x';\tau)$ is the heat kernel for $\beta=1$.

 The Maxwell field one-loop action in an arbitrary curved background  is:
  \begin{equation}
  \Gamma_{\rm M} ^{[1]}=-\frac {i}{2}\int_0^\infty  \frac{d\tau}{\tau}\int d^4x h^{\beta } _{\mu ,}\hspace{0.1 mm}^{\mu }(x,x;\tau)
  \label{mylady}
  \end{equation}
 with the gauge dependence  according to (\ref{Kermes}):
   \begin{equation}
  \beta \frac{\partial}{\partial \beta}   \Gamma_{\rm M} ^{[1]}=\frac i2 \int_0^\infty d\tau \int d^4x \frac{\partial}{\partial \tau}h(x, x;\frac 1\beta \tau).
  \label{chabrol}
  \end{equation}
  Also, the ghost action is here:
   \begin{equation}
  \Gamma ^{[1]}_{\rm M, gh}=i\int _0^\infty \frac{d\tau}{\tau}\int d^4xh (x, x;\frac{1}{\sqrt \beta}\tau)
  \label{karabagh}
  \end{equation}
  with:
   \begin{equation}
  \beta \frac{\partial}{\partial \beta} \Gamma^{[1]}_{\rm M, gh}=-\frac i2\int d^4x \int _0^\infty d\tau \frac{\partial}{\partial \tau}h (x, x;\frac{1}{\sqrt \beta}\tau)
  \label{shirvan}
  \end{equation}
  that formally cancels with (\ref{chabrol}) by a rescaling of $\tau$ but where the cancellation is upset by divergent terms. The determination of these divergent terms is carried out by (\ref{malacca}) and (\ref{aurora}) in the same way as for (\ref{demongeot}) and (\ref{mylene}), and one gets  from (\ref{chabrol}):
   \begin{equation}
    \beta \frac{\partial }{\partial \beta}\Gamma _{\rm M} ^{[1]}\simeq \frac 12 \frac{1}{16\pi^2}(-\frac{\beta^2}{\tau^2}+\frac 16R\frac{i\beta}{\tau})\mid _{\tau\simeq 0}
  \label{lemuria}
  \end{equation}
  and from (\ref{shirvan}):
  \begin{equation}
   \beta \frac{\partial}{\partial \beta} \Gamma^{[1]}_{\rm M, gh}\simeq -\frac 12 \frac{1}{16\pi^2}(-\frac{\beta}{\tau^2}+\frac 16R\frac{i\sqrt \beta}{\tau})\mid _{\tau\simeq 0}
  \label{atlantis}
  \end{equation}
  that do not cancel  at general values of $\beta$.

  \subsection{Gauge dependence at order $\kappa^2$ of Maxwell-Einstein theory}
 
In the Maxwell action $S_M$   a background field ${\cal A}_\mu$ is introduced,  with the corresponding field strength ${\cal F}_{\mu \nu}=\partial _\mu{\cal A}_\nu- \partial _\nu{\cal A}_\mu$,  with:
  \begin{equation}
  {\cal F}_{\mu \nu;}\hspace{0.1 mm}^\mu=0
  \label{mabeltrask}
  \end{equation}
  and the field  $A_\mu$ is split according to:
 \begin{equation}
 A_\mu \rightarrow {\cal A}_\mu +A_\mu
 \label{stocksplit}
 \end{equation}
 with $A_\mu$ still the quantum field.    The two-point correlation function is:
 \begin{equation}
 <A_\mu(x)A_{\lambda'}(x')>=\int _0^\infty   d\tau \frac{1}{^4 \sqrt{-g}}h^\beta_{\mu  \lambda'}(x,x';\tau)\frac{1}{ ^4 \sqrt{-g'}}.
 \label{elderberry}
 \end{equation}

The identity (\ref{CBH}) makes it straightforward to carry out a perturbation expansion of the effective action in the proper-time representation. At second order in $\kappa$ there is   a two-point function  term  by (\ref{anaximander}):
 \begin{eqnarray}&&
\Gamma ^{[1]}_{{\bf EM}, I}= \frac i2\kappa ^2\int d^4x\int d^4x'\int _0^\infty \tau d\tau\int _0^1 dth^\alpha_{\mu \nu, \xi'\eta'}(x, x';t\tau)
 (h^\beta _{\rho \delta'}(x,x';(1-t)\tau)_{;\lambda;\gamma'}
  \nonumber\\&&
  -h^\beta _{\rho \gamma'}(x,x';(1-t)\tau)_{;\lambda;\delta'}-h^\beta _{\lambda \beta'}(x,x';(1-t)\tau)_{;\rho;\alpha'}+h^\beta _{\lambda \gamma'}(x,x';(1-t)\tau)_{;\rho;\delta'})
  \nonumber\\&&
 (g^{\nu \lambda}{\cal F}^{\mu \rho}-\frac 14g^{\mu \nu}{\cal F}^{\lambda \rho})(x)(g^{\eta'\gamma'}{\cal F}^{\xi'\delta'}-\frac 14 g^{\xi'\eta'}{\cal F}^{\gamma'\delta'})(x' ).
 \label{mezquita}
 \end{eqnarray}
  This expression  has a quadratic divergence at $\alpha=1$ by (\ref{cornelia}):
  \begin{eqnarray}&&
\Gamma ^{[1]}_{{\bf EM}, I}\simeq -\frac 32\frac{i}{16\pi^2}\frac 1\tau \mid_{\tau\simeq 0} \kappa^2\int d^4 x\sqrt{-g}{\cal F}^{\mu \nu}{\cal F}_{\mu \nu}(x).
  \label{princetroy}
  \end{eqnarray}
    Also there is to this order a tadpole term by (\ref{mattgroening}):
 \begin{eqnarray}&&
\Gamma ^{[1]}_{{\bf EM},II}=  \kappa^2\int d^4x\int_0^\infty d\tau h^\alpha_{\omega \tau, \iota \sigma}(x, x;\tau) (\frac{1}{8}g^{\omega \tau}({\cal  F}^\iota \hspace{0.1  mm}_\lambda {\cal F}^{\sigma \lambda}-\frac 14g^{\iota \sigma} {\cal F}^{\mu \nu} {\cal F}_{\mu \nu})
 \nonumber\\&&
 +\frac{1}{16}g^{\omega \iota}g^{\sigma \tau} {\cal F}^{\mu \nu} {\cal F}_{\mu \nu}
 +\frac 18g^{\omega \tau} {\cal F}^\iota \hspace{0.1  mm}_\lambda {\cal F}^{\sigma \lambda}
 -\frac 14 {\cal F}^{\omega \iota} {\cal F}^{\tau \sigma}-\frac 12 g^{\sigma \omega} {\cal F}^\iota \hspace{0.1  mm}_\lambda {\cal F}^{\tau \lambda})
\label{tadpole}
 \end{eqnarray}
 with the quadratically divergent part  at $\alpha=1$  by (\ref{cordova}):
   \begin{equation}
 \Gamma ^{[1]}_{{\bf EM}, II}\simeq   \frac 34\frac{i}{16\pi^2}\frac 1\tau \mid_{\tau\simeq 0}\kappa^2\int d^4x\sqrt{-g}{\cal F}^{\mu \nu} {\cal F}_{\mu \nu}(x).
 \label{ireneadler}
 \end{equation}

 (\ref{mezquita}) does not depend on the gauge parameter $\beta$, and the dependence on $\alpha$ is in the lowest approximation  found from (\ref{tinaogmarina}):
 \begin{eqnarray}&&
 \alpha \frac{\partial}{\partial \alpha}\Gamma^{[1]}_{{\bf EM}, I}\simeq -\frac 12\frac 1\alpha \kappa^2\int d^4x \int d^4x'\int d^4x''\int _0^\infty \tau^2 d\tau \int _0^1 dt du dv \delta (1-t-u-v)
 \nonumber\\&&
  h_{{\rm gh}\mu \upsilon''}(x, x'';t\frac 1\alpha \tau)g^{\upsilon''\omega''}(x'')
  (h^\alpha _{\sigma ''\omega'', \xi' \eta'}(x'', x'; u\tau)_;\hspace{0.1 mm}^{\sigma ''}
 -\frac 12h^\alpha _{\sigma ''}\hspace{0.1 mm}^{\sigma''}\hspace{-1 mm}_{, \xi' \eta'}(x'', x'; u\tau)_{;\omega''})
 \nonumber\\&&
  (h^\beta _{\rho \delta'}(x,x'; v\tau)_{;\lambda;}\hspace{0.1 mm}^\lambda \hspace{0.1 mm}_{;\gamma'}
  -h ^\beta_{\rho \gamma'}(x,x'; v\tau)_{;\lambda;}\hspace{0.1 mm}^\lambda \hspace{0.1 mm}_{;\delta'}-h^\beta _{\lambda \delta'}(x,x'; v\tau)_{;\rho;}\hspace{0.1 mm}^\lambda \hspace{0.1 mm}_{;\gamma'}+h^\beta _{\lambda \gamma'}(x,x'; v\tau)_{;\rho;}\hspace{0.1 mm}^\lambda \hspace{0.1 mm}_{;\delta'})
  \nonumber\\&&
  {\cal F}^{\mu \rho}(x)
 ({\cal F}^{\xi'\delta'}g^{\eta'\gamma'}-\frac 14 g^{\xi'\eta'}{\cal F}^{\gamma'\delta'})(x' ).
 \label{generalife}
 \end{eqnarray}
  Using  (\ref{pontus}) in connection with:
  \begin{equation}
 h _{\lambda \beta'}(x,x'; \tau)_{;}\hspace{0.1 mm}^\lambda \hspace{0.1 mm}_{;\rho;\alpha'}
 -h _{\lambda \alpha'}(x,x';\tau)_{;}\hspace{0.1 mm}^\lambda \hspace{0.1 mm}_{;\rho;\beta'} =0,
 \label{sunflower}
 \end{equation}
following from  (\ref{monpetit}), one gets from (\ref{generalife}) two terms:    
\begin{eqnarray}&&
  \alpha \frac{\partial}{\partial \alpha}\Gamma^{[1]}_{{\bf EM}, I}\rightarrow   \frac 12\frac 1\alpha i\kappa^2\int d^4x \int d^4x'\int d^4x''\int _0^\infty  d\tau \frac{\partial}{\partial \tau}\tau^2\int _0^1 dt du dv \delta (1-t-u-v)
 \nonumber\\&&
  h_{{\rm gh}\mu \upsilon''}(x, x'';t\frac 1\alpha \tau)g^{\upsilon''\omega''}(x'')
  (h^\alpha _{\sigma ''\omega'', \xi' \eta'}(x'', x'; u\tau)_;\hspace{0.1 mm}^{\sigma ''}
 -\frac 12h^\alpha _{\sigma ''}\hspace{0.1 mm}^{\sigma''}\hspace{-1 mm}_{, \xi' \eta'}(x'', x'; u\tau)_{;\omega''})
 \nonumber\\&&
  (h^\beta _{\rho \delta'}(x,x'; v\tau)_{;\gamma'}
  -h ^\beta_{\rho \gamma'}(x,x'; v\tau)_{;\delta'})
  {\cal F}^{\mu \rho}(x)
 ({\cal F}^{\xi'\delta'}g^{\eta'\gamma'}-\frac 14 g^{\xi'\eta'}{\cal F}^{\gamma'\delta'})(x' )
    \label{perritocaliente}
    \end{eqnarray}
    and also, using the Bianchi identity and the field equation of the background gauge field:
        \begin{eqnarray}&&
  \alpha \frac{\partial}{\partial \alpha}\Gamma^{[1]}_{{\bf EM}, I}\rightarrow      -\frac 12\frac 1\alpha i\kappa^2\int d^4x \int d^4x'\int _0^\infty \tau d\tau \int _0^1 dt  
  h_{{\rm gh}\mu \upsilon'}(x, x';t\frac 1\alpha \tau)_{;\gamma}g^{\upsilon'\omega'}(x')
   \nonumber\\&&
  (h^\alpha _{\sigma '\omega', \xi \eta}(x', x; (1-t)\tau)_;\hspace{0.1 mm}^{\sigma '}
 -\frac 12h^\alpha _{\sigma '}\hspace{0.1 mm}^{\sigma'}\hspace{-1 mm}_{, \xi' \eta'}(x', x; (1-t)\tau)_{;\omega'})
 \nonumber\\&&
 g_{\rho \delta}  {\cal F}^{\mu \rho}(x)
 ({\cal F}^{\xi\delta}g^{\eta\gamma}-{\cal F}^{\xi\gamma}g^{\eta\delta}-\frac 12 g^{\xi\eta}{\cal F}^{\gamma\delta})(x )
 \nonumber\\&&
 -\frac 14\frac 1\alpha i\kappa^2\int d^4x \int d^4x'\int _0^\infty \tau d\tau \int _0^1 dt  
  h_{{\rm gh}\mu \upsilon'}(x, x';t\frac 1\alpha \tau)g^{\upsilon'\omega'}(x')
   \nonumber\\&&
  (h^\alpha _{\sigma '\omega', \xi \eta}(x', x; (1-t)\tau)_;\hspace{0.1 mm}^{\sigma '}
 -\frac 12h^\alpha _{\sigma '}\hspace{0.1 mm}^{\sigma'}\hspace{-1 mm}_{, \xi' \eta'}(x', x; (1-t)\tau)_{;\omega'})
 \nonumber\\&&
 ({\cal F}^{\xi \rho}{\cal F}^\eta \hspace{0.1 mm}_\rho-\frac 14 g^{\xi\eta}{\cal F}^{\rho \lambda}{\cal F}_{\rho \lambda})_{;}\hspace{0.1 mm}^{\mu}(x).
    \label{corazon}
    \end{eqnarray}

    Turning  to (\ref{tadpole}) one gets by (\ref{tinaogmarina}):
    \begin{eqnarray}&&
  \alpha \frac{\partial}{\partial \alpha}\Gamma^{[1]}_{{\bf EM},II}\simeq   -\frac 1\alpha i\kappa^2\int d^4 xd^4x'\int _0^\infty \tau d\tau\int _0^1 dt h_{{\rm gh}, \omega }\hspace{0.1 mm}^{\rho'}(x, x';t\frac 1\alpha \tau)_{;\upsilon}
    \nonumber\\&&
    (h^\alpha _{\sigma '\rho', \xi \eta}(x', x; (1-t)\tau)_;\hspace{0.1 mm}^{\sigma '}
 -\frac 12h^\alpha _{\sigma '}\hspace{0.1 mm}^{\sigma'}\hspace{-1 mm}_{, \xi \eta}(x', x; (1-t)\tau)_{;\rho'})
 \nonumber\\&&
 (\frac 14 g^{\omega \upsilon}({\cal F}^\xi \hspace{0.1 mm}_\lambda {\cal F}^{\eta \lambda}-\frac 14 g^{\xi \eta}{\cal F}^{\lambda \epsilon}{\cal F}_{\lambda \epsilon})+\frac 18g^{\omega \xi}g^{\upsilon \eta}{\cal F}^{\lambda \epsilon}{\cal F}_{\lambda \epsilon}+\frac 14 g^{\xi \eta}{\cal F}^\omega \hspace{0.1 mm}_\lambda {\cal F}^{\upsilon \lambda}-\frac 12 {\cal F}^{\omega \xi}{\cal F}^{\upsilon \eta}
 \nonumber\\&&
 -\frac 12g^{\omega \xi}{\cal F}^{\upsilon}\hspace{0.1 mm}_\lambda{\cal F}^{\eta \lambda} -\frac 12g^{\upsilon \xi}{\cal F}^{\omega}\hspace{0.1 mm}_\lambda{\cal F}^{\eta \lambda})(x)
  \label{foxglove}
 \end{eqnarray}
 and adding  (\ref{corazon}) and (\ref{foxglove}) to  (\ref{lamborghini}) one finally gets:
 \begin{eqnarray}&&
  \alpha \frac{\partial}{\partial \alpha}\Gamma^{[1]}\rightarrow  i\frac 1\alpha\int d^4x\int d^4x' \int_0^\infty  \tau d\tau\int _0^1 dth_{{\rm gh},}\hspace{0.1 mm}^{\mu \sigma'}(x, x';t\frac 1\alpha \tau)
  ( {\cal G}^{\omega '\delta'}-{\cal T}^{\omega '\delta'})(x')
    \nonumber\\&&
  ( h^\alpha_{\omega'\sigma',\mu \nu}(x', x;(1-t)\tau)_{;\delta'}\hspace{0.1 mm}^{;\nu}-\frac 12  h^\alpha_{\omega'\sigma',\nu}\hspace{0.1 mm}^{ \nu}(x', x;(1-t)\tau)_{;\delta'}\hspace{0.1 mm}_{;\mu}
   \nonumber\\&&
   -\frac 12(h^\alpha_{\omega'\delta',\mu \nu}(x', x;(1-t)\tau)_{;\sigma'}\hspace{0.1 mm}^{;\nu}
  -\frac 12  h^\alpha_{\omega'\delta',\nu}\hspace{0.1 mm}^{ \nu}(x', x;(1-t)\tau)_{;\sigma'}\hspace{0.1 mm}_{;\mu}))
 \label{avonlea}
 \end{eqnarray}
 with ${\cal T}^{\mu \nu}$ the background gauge field energy-momentum tensor,
 i. e. the Einstein tensor ${\cal G}^{\mu \nu}$ in (\ref{lamborghini}) gets the additional term $-{\cal T}^{\mu\nu}$  when the gravitational field is coupled to an Abelian gauge field.  It has been verified that this conclusion also holds for a non-Abelian gauge field, but the proof is not included in this article.  The quadratic divergence of (\ref{avonlea}) is still given by (\ref{kilkenney}) because the trace of the Maxwell field energy-momentum tensor vanishes in four dimensions.

 Finding the quadratic divergence of (\ref{perritocaliente}) one uses the Ward identity (\ref{kalmar}), keeping only the first term on the right hand side, in connection with (\ref{semigroup}), obtaining by a partial integration:
  \begin{eqnarray}&&
  \frac 12 \frac 1\alpha  i\kappa^2\int d^4x \int d^4x'\int _0^\infty  d\tau \frac{\partial}{\partial \tau}\tau^2\int _0^1 tdt h_{{\rm gh}\mu \xi'}(x, x';t\frac 1\alpha \tau)
 \nonumber\\&&
  (h^\beta _{\rho \delta'}(x,x'; (1-t)\tau)_{;\gamma';\eta'}
  -h ^\beta_{\rho \gamma'}(x,x'; (1-t)\tau)_{;\delta';\eta'})
   \nonumber\\&&
  {\cal F}^{\mu \rho}(x)
 ({\cal F}^{\xi'\delta'}g^{\eta'\gamma'}+{\cal F}^{\eta'\delta'}g^{\xi'\gamma'}-\frac 12 g^{\xi'\eta'}{\cal F}^{\gamma'\delta'})(x' ).
     \label{belinda}
    \end{eqnarray}
(\ref{belinda}) is evaluated by (\ref{askemose}) and contains the quadratic divergence:   
\begin{equation}
    -\frac 38\alpha \frac{i}{16\pi^2}\frac 1\tau\mid _{\tau \rightarrow 0}\kappa^2\int d^4x \sqrt{-g} {\cal F}^{\mu\nu} {\cal F}_{\mu\nu}(x).
  \label{stenager}
  \end{equation}
  Inserting (\ref{stenager}) in (\ref{perritocaliente}), integrating and adding  (\ref{princetroy}) and (\ref{ireneadler})  one obtains: 
  \begin{equation}
 -\frac 38 (1+\alpha)\frac{i}{16\pi^2}
  \frac 1\tau\mid _{\tau \rightarrow 0} \kappa^2 \int d^4x \sqrt{-g} {\cal F}^{\mu\nu} {\cal F}_{\mu\nu}(x)
  \label{effendi}
  \end{equation}
  in agreement with Toms \cite{Toms}, \cite{TomsII} when the proper time $\tau$ is converted to a temperature $T$ by taking it imaginary.   However, it should be kept in mind that the expression (\ref{perritocaliente}), which is responsible for the gauge parameter dependence of (\ref{effendi}), contains a total derivative in the proper time integral and thus is in the same category as (\ref{bergerac}) and (\ref{shehrezade}), where the gauge dependence was not removed by the Vilkovisky construction. This is the case also for (\ref{perritocaliente}) as discussed in the following section.

\subsection{Vilkovisky's construction in Maxwell-Einstein theory}

In Maxwell-Einstein theory  (\ref{mikkelmax}) is replaced by:
\begin{equation}
S_{, h_{\mu \nu}}=-\frac {1}{\kappa}({\cal G}^{\mu\nu}-{\cal T}^{\mu \nu}).
\label{makkelmix}
\end{equation}
In  (\ref{starman}) and (\ref{prokofiev}) one thus has to carry out the replacement ${\cal G}^{\mu \nu}\rightarrow  {\cal G}^{\mu\nu}-{\cal T}^{\mu \nu}$, and adding (\ref{starman}) after this replacement  to (\ref{avonlea})  with the parameter $k$ fixed at $-1$  one removes the dependence on the gauge parameter $\alpha$.
  Thus the Vilkovisky construction of quantum gravity is sufficient to remove the gauge dependence also of the full Einstein-Maxwell system in lowest order, without additional modifications. The vanishing of the quadratic divergence of (\ref{prokofiev}) persists after the replacement, and the gauge parameter dependence of (\ref{effendi}) is not eliminated through the Vilkovisky construction.

  For the Maxwell field one has:
\begin{eqnarray}&&
R^{A_\mu(x)}\hspace{0.1 mm}_{c(y)}\simeq \delta(x, y)_{,\mu},
\label{shostakovich}
\end{eqnarray}
and the field metric is:
\begin{equation}
G_{A_\mu(x)A_\nu(y)}=\sqrt{-g}g^{\mu \nu}(x)\delta (x, y).
\label{oddeidem}
\end{equation}
The projection operator $\Pi ^m\hspace{0.1 mm}_n$ corresponding to the Maxwell field  is  in lowest order, cp. (\ref{carolina}): 
 \begin{equation}
 \Pi_{\rho \mu '} (x,x')=\Pi _{A^{\rho}{x}A^{\mu'}(x')}=g_{\rho \alpha'}\delta (x, x')-i^4\sqrt{-g}\frac{1}{\sqrt \beta}<c(x)_{,\rho}\bar c(x')_{,\mu'}>\hspace{0.1 mm}^4\sqrt{-g'}
 \label{deichgraf}
 \end{equation}
 with the ghost propagator expressed in terms of the scalar heat kernel defined in (\ref{lufthanna}):
  \begin{equation}
 <c(x)\bar c(x'))>=\frac{1}{\sqrt \beta} \int_0^{\infty} d\tau \frac{1}{^4 \sqrt{-g}}h(x,x';\frac 1\beta \tau)\frac{1}{ ^4 \sqrt{-g'}}.
 \label{embla}
 \end{equation}
This operator  projects from the photon propagator in an arbitrary gauge the transverse photon propagator. 
 From (\ref{Maxwell}) and (\ref{Maxbreak}) follows for the two-point function:
    \begin{equation}
 <F_{\lambda \rho;}\hspace{0.1 mm}^\lambda(x)A_{\mu '}(x')>
 = i\frac{1}{^4\sqrt{-g}}\Pi_{\rho \mu '} (x,x')\frac{1}{^4\sqrt{-g'}}.
  \label{digegreve}
 \end{equation}
 
The Christoffel connection components are:
\begin{eqnarray}&&
\Gamma ^{g_{\rho \sigma}(x)}\hspace{0.1 mm}_{A_\mu(y)A_\nu(z)}=\kappa^2\delta _{(\sigma \omega)}\hspace{0.1 mm}^{(\mu \nu)}\delta (x, y)\delta (y, z).
\label{DeWitt}
\end{eqnarray}
From (\ref{vildtfoged}) one thus gets the connection coupling term in the effective action:
\begin{equation}
\frac {1}{2}\int d^4x\sqrt{-g}({\cal G}^{\mu\nu}-{\cal T}^{\mu \nu})
(x)
<A_\mu(x)A_\nu(x)>
\label{maserati}
\end{equation}
with a transverse photon propagator, and with the quadratic divergence:
\begin{equation}
\frac 32 \frac{i}{16\pi^2}\frac 1\tau \mid_{\tau \simeq 0}\int d^4x\sqrt{-g}R
\label{colinfirth}
\end{equation}
with $R$ the scalar curvature and with no contribution from the background gauge field.

\subsection{General gauge fixing}

The gauge breaking action (\ref{GB}) can be generalized to:
\begin{equation}
S_{GB}=-\frac 12\frac 1\alpha\int d^4x\sqrt{-g}g^{\mu \nu}\chi_\mu \chi_\nu
\label{aronvullerslev}
\end{equation}
with:
\begin{equation}
\chi _\mu =h_{\mu \nu ;}\hspace{0.1 mm}^\nu-\frac 12 g^{\nu \sigma}h_{\nu \sigma;\mu}+\kappa (\omega_1 {\cal A}_\mu A^\lambda \hspace{0.1 mm}_{;\lambda}+\omega _2{\cal F}_{\lambda \mu}A^\lambda)
\label{samsonvullerslev}
\end{equation}
where $\omega_1$ and $\omega_2$ are new gauge parameters. This gives rise to new couplings:
\begin{equation}
- \frac 1\alpha \kappa\int d^4x\sqrt{-g} g^{\mu \nu}(h_{\mu \lambda;}\hspace{0.1 mm}^\lambda-\frac 12 h^\lambda\hspace{0.1 mm}_{\lambda;\mu})(\omega_1 {\cal A}_\nu A^\rho \hspace{0.1 mm}_{;\rho}+\omega _2{\cal F}_{\rho \nu}A^\rho)
\label{tjalfekristoffer}
\end{equation}
and:
\begin{equation}
-\frac 12\frac 1\alpha\kappa^2\int d^4x\sqrt{-g}g^{\mu \nu}(\omega_1 {\cal A}_\mu A^\lambda \hspace{0.1 mm}_{;\lambda}+\omega _2{\cal F}_{\lambda \mu}A^\lambda)(\omega_1 {\cal A}_\nu A^\rho \hspace{0.1 mm}_{;\rho}+\omega _2{\cal F}_{\rho \nu}A^\rho).
\label{kajuskristoffer}
\end{equation}
 The corresponding ghost action replacing (\ref{ghost}) and (\ref{Maxghost}) is, keeping only terms relevant at one-loop order:
\begin{eqnarray}&&
S_{FP}=\frac{1}{\sqrt{\alpha}}\int d^4x\sqrt{-g}\bar \xi ^\mu \bigg(\xi_{\mu;\nu;}\hspace{0.1 mm}^\nu+R_{\mu \nu}\xi^\nu\ 
 +\kappa (\omega_1{\cal A}_\mu c_{,}\hspace{0.1 mm}^\kappa\hspace{0.1 mm}_{;\kappa}+\omega_2{\cal F}_{\lambda \mu}c_{;}\hspace{0.1  mm}^\lambda)
 \nonumber\\&&
 +\kappa^2(\omega_1{\cal A}_\mu((\xi^\lambda{\cal A}_{\lambda})_{,\rho}+\xi^\lambda {\cal F}_{\lambda\rho})_{;}\hspace{0.1 mm}^\rho+\omega_2{\cal F}_{\rho \mu}((\xi^\lambda\hspace{0.1 mm}{\cal A}_{\lambda})_{;}\hspace{0.1  mm}^{\rho}+\xi^\lambda {\cal F}_\lambda \hspace{0.1 mm}^{\rho}))\bigg)
\nonumber\\&&
+\frac{1}{\sqrt{\beta}}\int d^4x\sqrt{-g}\bar c(c_{,\mu}+\kappa((\xi^\lambda{\cal A}_{\lambda})_{,\mu}+\xi^\lambda {\cal F}_{ \lambda\mu}))_{;}\hspace{0.1 mm}^\mu. 
\label{dartaurus}
\end{eqnarray}

A new one-loop term in the effective action of order  $\kappa^2$  is by (\ref{tjalfekristoffer}):
\begin{eqnarray}&&
\frac i2\frac {1}{\alpha ^2}\kappa ^2 \int d^4x\sqrt{-g}  \int d^4x'\sqrt{-g'} 
<(h_{\mu \nu;}\hspace{0.1 mm}^{\nu}-\frac 12 h^{\nu}\hspace{0.1 mm}_{\nu;\mu})(x)
(h_{\lambda'\rho';}\hspace{0.1 mm}^{\rho'}-\frac 12 h^{\rho'}\hspace{0.1 mm}_{\rho';\lambda'})(x')>
\nonumber\\&&
<(\omega_1{\cal A}^{\mu}A^{\sigma}\hspace{0.1 mm}_{;\sigma}+\omega_2{\cal F}^{\sigma\mu}A_{\sigma})(x)(\omega_1{\cal A}^{\lambda'}A^{\gamma'}\hspace{0.1 mm}_{;\gamma'}+\omega_2{\cal F}^{\gamma'\lambda'}A_{\gamma'})(x')>.
\label{kildegaard}
\end{eqnarray}
 (\ref{kildegaard}) is expressed in the  proper time representation and the Ward identity (\ref{kalmar}) is applied, disregarding the last term containing the Einstein tensor. Then (\ref{kildegaard})  is:
\begin{eqnarray}&&
-\frac i2\frac {1}{\alpha ^2} \kappa ^2\int d^4x  \int d^4x'
\int_0^\infty \tau d\tau 
\int_0^1 dt(h_{{\rm gh}, \lambda'\mu}(x', x;\frac 1\alpha t\tau)_{;\rho';}\hspace{0.1 mm}^{\rho'}
+R_{\lambda'\rho'}h_{{\rm gh},}\hspace{0.1 mm}^{ \rho'}\hspace{0.1 mm}_{\mu}(x', x;\frac 1\alpha t\tau))
\nonumber\\&&
(\omega_1^2{\cal A}^{\mu}(x){\cal A}^{\lambda'}(x')h^{\beta, \sigma\gamma'}(x, x';(1-t)\tau)_{;\sigma;\gamma'}+\omega_2^2{\cal F}^{\sigma\mu}(x){\cal F}^{\gamma'\lambda'}(x')
h^{\beta}\hspace{0.1 mm}_{ \sigma\gamma'}(x, x';(1-t)\tau)
\nonumber\\&&
+2\omega _1\omega_2{\cal A}^{\mu}(x){\cal F}^{\gamma'\lambda'}(x')h^{\beta}\hspace{0.1 mm}_{ \sigma\gamma'}(x, x';(1-t)\tau)_{;}\hspace{0.1 mm}^{\sigma})
\label{lombjerge}
\end{eqnarray}
 containing two terms by (\ref{lindberg})-(\ref{anna}):
\begin{eqnarray}&&
-\frac 12\frac {1}{\alpha } \kappa ^2\int d^4x  \int d^4x'
\int_0^\infty d\tau \frac{\partial}{\partial \tau}\tau \int_0^1 dth_{{\rm gh}, \lambda'\mu}(x', x;\frac 1\alpha t\tau)
\nonumber\\&&
(\omega_1^2{\cal A}^{\mu}(x){\cal A}^{\lambda'}(x')h^{\beta, \sigma\gamma'}(x, x';(1-t)\tau)_{;\sigma;\gamma'}+\omega_2^2{\cal F}^{\sigma\mu}(x){\cal F}^{\gamma'\lambda'}(x')
h^{\beta}\hspace{0.1 mm}_{ \sigma\gamma'}(x, x';(1-t)\tau)
\nonumber\\&&
+2\omega _1\omega_2{\cal A}^{\mu}(x){\cal F}^{\gamma'\lambda'}(x')h^{\beta}\hspace{0.1 mm}_{ \sigma\gamma'}(x, x';(1-t)\tau)_{;}\hspace{0.1 mm}^{\sigma})
\label{birknakke}
\end{eqnarray}
and:
\begin{equation}
\frac 12\frac 1\alpha\kappa ^2 \int d^4x\sqrt{-g}g^{\mu \nu}<(\omega_1 {\cal A}_\mu A^\lambda \hspace{0.1 mm}_{;\lambda}+\omega_2{\cal F}_{\lambda \mu}A^\lambda)(x)(\omega_1{\cal A}_\nu A^\rho \hspace{0.1 mm}_{;\rho}+\omega _2{\cal F}_{\rho \nu}A^\rho)(x)>.
\label{kajusdar}
\end{equation}
(\ref{birknakke})  contains an integral of a total derivative in the proper-time variable $\tau$ in the same way as (\ref{perritocaliente}) and only has quartical and quadratically divergent terms, and it should be disregarded  formally; its gauge parameter dependence is not removed through the Vilkovisky construction.  
Also we get from (\ref{kajuskristoffer}):
\begin{equation}
-\frac 12\frac {1}{\alpha}\kappa^2\int d^4x\sqrt{-g}g^{\mu \nu}<(\omega_1 {\cal A}_\mu A^\lambda \hspace{0.1 mm}_{;\lambda}+\omega_2{\cal F}_{\lambda \mu}A^\lambda)(x)(\omega_1{\cal A}_\nu A^\rho \hspace{0.1 mm}_{;\rho}+\omega _2{\cal F}_{\rho \nu}A^\rho)(x)>
\label{kristofferryslinge}
\end{equation}
that cancels (\ref{kajusdar}).

Also there is a cross term from (\ref{anaximander}) and (\ref{tjalfekristoffer}):
\begin{eqnarray}&&
-i\frac {\omega_2}{\alpha }\kappa ^2\int d^4x\sqrt{-g}\int d^4x'\sqrt{-g'}<(h_{\mu \nu;}\hspace{0.1 mm}^{\nu}-\frac 12 h^{\nu}\hspace{0.1 mm}_{\nu;\mu})(x)h_{\lambda'\rho'}(x')>
\nonumber\\&&
{\cal F}^{\sigma\mu}(x)(g^{\lambda'\gamma'}{\cal F}^{\rho'\delta'}-\frac 14 g ^{\lambda'\rho'}{\cal F}^{\gamma'\delta'})(x')<A_{\sigma}(x)F_{\gamma'\delta'}(x')>.
\label{primusbirk}
\end{eqnarray}
 Here  the proper time representation again is used, combined with the Ward identity (\ref{kalmar}) with the term containing the Einstein tensor disregarded, with the result obtained by partial integration:
 \begin{eqnarray}&&
-i\frac{\omega_2}{\alpha}\kappa ^2 \int d^4x\int d^4x'\int _0^\infty \tau d\tau \int _0^1 dth_{{\rm gh},\rho'\mu}(x', x;\frac 1\alpha t\tau) {\cal F}^{\sigma\mu}(x){\cal F}^{\rho'\delta'}(x')
 \nonumber\\&&
(h^{\beta}\hspace{0.1 mm}_{ \sigma}\hspace{0.1 mm}_{\delta'}(x, x';(1-t)\tau)_{;}\hspace{0.1 mm}^{\gamma'}\hspace{0.1 mm}_{;\gamma'}-h^{\beta}\hspace{0.1 mm}_\sigma \hspace{0.1 mm}^{ \gamma'}(x, x';(1-t)\tau)_{;\delta';\gamma'})
 \label{kristofferdannis}
 \end{eqnarray}
with two terms by (\ref{pontus}) and (\ref{frits}) and the Ward identity (\ref{monpetit}):
 \begin{eqnarray}&&
 -\frac{\omega_2}{\alpha}\kappa ^2 \int d^4x_1\int d^4x_2\int _0^\infty  d\tau \frac{\partial}{\partial \tau}\tau\int _0^1 dth_{{\rm gh}, \rho'\mu}(x', x;\frac 1\alpha t\tau) {\cal F}^{\sigma\mu}(x){\cal F}^{\rho'\delta'}(x')
 \nonumber\\&&
h^{\beta}_{\sigma\delta'}(x, x';(1-t)\tau)
\label{kristofferbroby}
 \end{eqnarray}
  and:
 \begin{eqnarray}&&
 \frac{\omega_2}{\alpha}\kappa^2\int d^4x\int _0^\infty d\tau h_{{\rm gh}, \rho\mu}(x,x;\frac 1\alpha \tau){\cal F}_{\sigma}\hspace{0.1 mm}^{\mu}(x){\cal F}^{\rho\sigma}(x)
 \nonumber\\&&
 -i\frac{\omega_2}{\alpha\beta}\kappa ^2 \int d^4x\int d^4x'\int _0^\infty \tau d\tau \int _0^1 dth_{{\rm gh}\rho'\mu}(x', x;\frac 1\alpha t\tau) {\cal F}^{\sigma\mu}(x){\cal F}^{\rho'\delta'}(x')
 \nonumber\\&&
h(x, x';\frac 1\beta(1-t)\tau)_{;\sigma;\delta'}.
 \label{juliechristie}
 \end{eqnarray}
 (\ref{kristofferbroby}) is again an expression like (\ref{perritocaliente}), containing a proper-time integral of a total differential. 
  
 From the ghost action (\ref{dartaurus}) one gets the new one-loop contributions to the effective action:
\begin{eqnarray}&&
i \frac{\omega_1}{\sqrt{\alpha \beta}}\kappa^2\int d^4x  {\cal A}^{\mu}(x)\int d^4x '\int _0^\infty \tau d\tau \int _0^1 dth(x, x';\frac{t\tau}{\sqrt{\beta}})_{;\rho;}\hspace{0.1 mm}^{\rho}\hspace{0.1 mm}_{;}\hspace{0.1 mm}^{\nu'}
\nonumber\\&&
({\cal A}^{\lambda'}(x')h_{{\rm gh}, \lambda'\mu}(x', x;\frac{(1-t)\tau}{\sqrt{\alpha}})_{;\nu'}+{\cal A}_{\nu';}\hspace{0.1 mm}^{\lambda'}(x)h_{{\rm gh}, \lambda'\mu}(x', x;\frac{(1-t)\tau}{\sqrt{\alpha}}))
\nonumber\\&&
-\frac{\omega_1}{\sqrt{\alpha}}\kappa^2\int d^4x{\cal A}_{\mu}(x)({\cal A}^{\lambda'}(x')
h_{{\rm gh}, \lambda'\mu}(x', x;\frac{(1-t)\tau}{\sqrt{\alpha}})_{;\nu'}
\nonumber\\&&
+
{\cal A}_{\nu';}\hspace{0.1 mm}^{\lambda'}(x')h_{{\rm gh}, \lambda'\mu}
(x', x;\frac{(1-t)\tau}{\sqrt{\alpha}}))_{;}\hspace{0.1 mm}^{\nu'}\mid_{x'\rightarrow x}
\nonumber\\&&
\simeq   \frac{\omega_1}{\sqrt{\alpha }}\kappa^2\int d^4x  {\cal A}^{\mu}(x)\int d^4x '\int _0^\infty  d\tau \frac{\partial}{\partial \tau}\tau\int_0^1 dth(x, x';\frac{t\tau}{\sqrt{\beta}})_{;}\hspace{0.1 mm}^{\nu'}
\nonumber\\&&
({\cal A}^{\lambda'}(x')h_{{\rm gh}, \lambda'\mu}(x', x;\frac{(1-t)\tau}{\sqrt{\alpha}})_{;\nu'}+{\cal A}_{\nu';}\hspace{0.1 mm}^{\lambda'}(x')h_{{\rm gh}, \lambda'\mu'}(x', x;\frac{(1-t)\tau}{\sqrt{\alpha}}))
\nonumber\\&&
\label{borgmesteren}
\end{eqnarray}
and:
\begin{eqnarray}&&
-i\frac{\omega _2}{\sqrt{\alpha \beta}}\kappa^2\int d^4x{\cal F}^{\rho\mu}(x)\int d^4x' {\cal A}_{\lambda'}(x')\int _0^\infty \tau d\tau\int _0^1 dth(x, x';\frac{1}{\sqrt \beta}t\tau)_{;\rho; \nu';}\hspace{0.1 mm}^{\nu'} 
\nonumber\\&&
h_{{\rm gh}}\hspace{0.1 mm}^{\lambda'\mu}(x', x;\frac{1}{\sqrt{\alpha}}(1-t)\tau)
\nonumber\\&&
+\frac{\omega_2}{\sqrt{\alpha}}\kappa^2\int d^4x{\cal F}^{\rho\mu}(x){\cal A}^{\lambda}(x)h_{{\rm gh}, \lambda \mu'}(x', x;\tau)_{;\rho}\mid _{x'\rightarrow x}
\nonumber\\&&
\simeq -\frac{\omega_2}{\sqrt{\alpha }}\kappa^2\int d^4x {\cal F}^{\rho \mu}(x) \int d^4x _2{\cal A}_{\lambda'}(x')\int_0^\infty d\tau\frac{\partial}{\partial \tau}\tau \int_0^1dth(x, x'; \frac{t\tau}{\sqrt{\ \beta}})_{;\rho}
\nonumber\\&&
h_{{\rm gh}}\hspace{0.1 mm}^{\lambda '\mu}(x', x;\frac{(1-t)\tau}{\sqrt{\alpha }})
\label{kristofferjuel}
\end{eqnarray}
by (\ref{lufthanna}), which again are of the same type as (\ref{perritocaliente}), with a proper time integral of a total derivative.
The  final effective action term arising from (\ref{dartaurus}) is:
 \begin{eqnarray}&&
  -\frac{\omega_2}{\sqrt \alpha}\kappa^2\int d^4x\int _0^\infty d\tau h_{{\rm gh}, \rho\mu}(x,x;\frac {1}{\sqrt{\alpha} }\tau){\cal F}_{\sigma}\hspace{0.1 mm}^{\mu}(x){\cal F}^{\rho\sigma}(x)
 \nonumber\\&&
 +i\frac{\omega_2}{\sqrt{\alpha\beta}}\kappa ^2 \int d^4x\int d^4x'\int _0^\infty \tau d\tau \int _0^1 dth_{{\rm gh}, \rho'\mu}(x', x;\frac {1}{\sqrt{\alpha}} t\tau) {\cal F}^{\sigma\mu}(x){\cal F}^{\rho'\delta'}(x')
 \nonumber\\&&
h(x, x';\frac {1}{\sqrt{\beta}}(1-t)\tau)_{;\sigma;\delta'}.
\label{edmundkristoffer}
\end{eqnarray}
In (\ref{juliechristie}) and (\ref{edmundkristoffer}) one can introduce new variables $\tau_1=t\tau$ and $\tau_2=(1-t)\tau$.  Then the two expressions cancel formally by  rescaling  of the variables $\tau$, $\tau_1$ and $\tau_2$. However, this argument is invalidated by quadratic divergences. This is similar to the imperfect cancellation between  (\ref{bergerac}) and (\ref{shehrezade}) and between (\ref{chabrol}) and (\ref{shirvan}).

The additional  terms in a general gauge are (\ref{kildegaard}),  (\ref{kristofferryslinge}),  (\ref{primusbirk}), (\ref{borgmesteren}), (\ref{kristofferjuel}) and (\ref{edmundkristoffer}), where the three first terms are modified into the sum of   (\ref{birknakke}), (\ref{kristofferbroby}) and   (\ref{juliechristie}). Formally the sum of these expressions  vanishes, but in the proper time representation with the proper time integrals regularized by a lower cut-off the vanishing of the sum is upset by quartic and quadratic divergences.    The values of  (\ref{birknakke}), (\ref{kristofferbroby}),   (\ref{borgmesteren}) and (\ref{kristofferjuel}) and the difference between (\ref{juliechristie}) and  (\ref{edmundkristoffer}) are all determined in Appendix A.

When using the Ward identity (\ref{kalmar}) we have disregarded the term on the right-hand side containing the Einstein tensor ${\cal G}^{\mu\nu}$. When including this term in the calculations reported in this section  and also terms of fourth order in $\kappa$ one generates new terms of the effective action containing one power of the combination ${\cal G}^{\mu\nu}-{\cal T}^{\mu \nu}$, with ${\cal T}^{\mu \nu}$ the background gauge field energy-momentum tensor.
The gauge parameter dependence of these terms is removed by the Vilkovisky construction in next-lowest order. The calculation is lengthy, but is important for the use of the Landau-DeWitt gauge condition; an outline  is given in Appendix B.

\section{Momentum space integrals}

The flat-space propagators in $D$ dimensions are:
\begin{eqnarray}&&
<h_{\mu \nu}(x)h_{\lambda'\rho'}(x')>=\int \frac {d^Dk}{(2\pi)^D}e^{ik(x-x')}\frac{-i}{k^2-i\epsilon}\bigg(g_{\mu \lambda'}g_{\nu \rho'}+g_{\nu \lambda'}g_{\mu \rho'}-\frac{2}{D-2}g_{\mu \nu}g_{\lambda'\rho'}
\nonumber\\&&
-(1-\alpha)\frac{1}{k^2}(k_\mu k_{\lambda'}g_{\nu \rho'}+k_\nu k_{\lambda'}g_{\mu \rho'}+k_\mu k_{\rho'}g_{\nu \lambda'}+k_\nu k_{\rho'}g_{\mu \lambda'})\bigg)
\label{kristofferholev}
\end{eqnarray}
 as well as:
\begin{equation}
<A_{\mu }(x)A_{\lambda'}(x')>=\int \frac {d^Dk}{(2\pi)^D}e^{ik(x-x')}\frac{-i}{k^2-i\epsilon}(\eta _{\mu \lambda'}
-(1-\beta)\frac{1}{k^2}k_\mu k_{\lambda'})
\label{kristoffertolvhave}
\end{equation}
and also:
\begin{equation}
<\xi_{\mu}(x)\bar \xi_{\nu'}(x')>=\sqrt \alpha\int \frac {d^Dk}{(2\pi)^D}e^{ik(x-x')}\frac{-i}{k^2-i\epsilon}\eta_{\mu \nu'}
\label{eskebrangstrup}
\end{equation}
with the Ward identity:
\begin{eqnarray}&&
<(\partial ^\mu h_{\mu \nu}-\frac 12 \partial _\nu h^\mu \hspace{0.1 mm}_\mu)(x)h_{\lambda'\rho'}(x')>
\nonumber\\&&
=\alpha \int \frac {d^Dk}{(2\pi)^D}e^{ik(x-x')}\frac{-i}{k^2-i\epsilon}i(k_{\lambda '}\eta _{\rho'\nu}+k_{\rho'}\eta _{\lambda'\nu})
\nonumber\\&&
=-\sqrt \alpha<(\partial _{\lambda'}\xi_{\rho'}+\partial _{\rho'}\xi_{\lambda'})(x')\xi_\nu(x)>
\label{kattekilde}
\end{eqnarray}
that has the immediate consequence:
\begin{equation}
<(\partial ^\mu h_{\mu \nu}-\frac 12 \partial _\nu h^\mu \hspace{0.1 mm}_\mu)(x)(\partial ^{\lambda '}h_{\lambda'\rho'}-\frac 12 \partial _{\rho'}h^{\lambda '} \hspace{0.1 mm}_{\lambda '}(x')>=-i\alpha\eta _{\nu \rho'} \delta(x-x').
\label{nissetid}
\end{equation}

At $D=4$ one gets from (\ref{mezquita}) when converting it to a momentum space integral: 
\begin{eqnarray}&&
\frac 12 (\frac{3}{D'}+(1-\frac{1}{D'})\alpha)\kappa^2\int \frac{d^4k}{(2\pi)^4}\frac{-i}{k^2-i\epsilon}
\int d^4x{\cal F}^{\mu \nu}(x){\cal F}_{\mu \nu}(x).
\label{bakewell}
\end{eqnarray}   Here was used:
\begin{equation}
\int \frac{d^Dk}{(2\pi)^D}f(k^2)k_\mu k_\nu =\frac{1}{D'}\int \frac{d^Dk}{(2\pi)^D}f(k^2)k^2
\label{simsalabim}
\end{equation}
with $f(k^2)$ arbitrary, where  possibly  $D'\neq D$ for quadratic divergences. In \cite{He} and \cite{Tang} it was argued that the value $D'=2$ should be used.  Also one gets from (\ref{tadpole})  for $D=4$:
\begin{equation}
-\frac 34\kappa ^2(1-\frac {2}{D'}(1-\alpha))
\int d^4x {\cal F}^{\mu \nu}{\cal F}_{\mu \nu}(x)\int \frac { d^4k}{(2\pi)^4}\frac{-i}{k^2-i\epsilon}.
\label{cyrusbirkum}
\end{equation}
The sum of  (\ref{bakewell}) and (\ref{cyrusbirkum}) at general $D'$ is:
 \begin{equation}
 (3-2\alpha)(\frac{1}{D'}-\frac 14) \kappa^2\int \frac{d^4k}{(2\pi)^4}\frac{-i}{k^2-i\epsilon} \int d^4x{\cal F}^{\mu \nu}(x){\cal F}_{\mu \nu}(x)
 \label{nebos}
 \end{equation}
 in agreement with Tang and Wu \cite{Tang}.

(\ref{nebos}) vanishes at $D'=4$,  and we have thus reproduced Pietrykowski's result \cite{Pietrykowski}, that the linear divergences of the effective action cancel for all values of $\alpha$. Taking instead $D'=2$ one gets from (\ref{nebos}):
\begin{equation}
\frac 14( 3+\alpha) \kappa^2\int \frac{d^4k}{(2\pi)^4}\frac{-i}{k^2-i\epsilon} \int d^4x{\cal F}^{\mu \nu}(x){\cal F}_{\mu \nu}(x).
\label{surumu}
\end{equation}
 in agreement with  He, Wang and Xianyu \cite{He}.

  (\ref{nebos}) is considered in connection with (\ref{starman}) converted to a momentum space integral with the replacement ${\cal G}^{\mu\nu}\rightarrow {\cal G}^{\mu\nu}-{\cal T}^{\mu \nu}$, where ${\cal T}^{\mu \nu}$ is the background gauge field energy-momentum tensor,  and  from which one gets at $D=4$:
 \begin{eqnarray}&&
 2\alpha (\frac{1}{D'}-\frac 14)\kappa^2\int \frac{d^4k}{(2\pi)^4}\frac{-i}{k^2-i\epsilon}
\int d^4x{\cal F}^{\mu\nu}{\cal F}_{\mu \nu}(x)
  \label{nebelwerfer}
 \end{eqnarray}
 where the sum indeed is independent of $\alpha$:
 \begin{equation}
 3(\frac{1}{D'}-\frac 14) \kappa^2 \int \frac{d^4k}{(2\pi)^4}\frac{-i}{k^2-i\epsilon} \int d^4x{\cal F}^{\mu \nu}(x){\cal F}_{\mu \nu}(x).
 \label{birkhahn}
 \end{equation}

 The contributions arising from the Vilkovisky connections (\ref{prokofiev}) and (\ref{maserati}) are, keeping in mind that the propagators are transverse:
\begin{eqnarray}&&
-(\frac{1}{D'}-\frac 14)\kappa^2\int \frac{d^4k}{(2\pi)^4}\frac{-i}{k^2-i\epsilon}
\int d^4x{\cal F}^{\mu \nu}{\cal F}_{\mu \nu}(x) 
\label{polovec}
\end{eqnarray}
and:
\begin{eqnarray}&&
(\frac{1}{D'}-\frac 14)\frac {\kappa^2}{4}\int \frac{d^4k}{(2\pi)^4}\frac{-i}{k^2-i\epsilon}\int d^4x{\cal F}^{\mu \nu}{\cal F}_{\mu \nu}(x)
\label{kuman}
\end{eqnarray}
 in agreement with Tang and Wu \cite{Tang}. The sum of (\ref{birkhahn}), (\ref{polovec}) and (\ref{kuman}) is:
 \begin{equation}
\frac{9}{4}(\frac{1}{D'}-\frac 14)\kappa^2 \int \frac{d^4k}{(2\pi)^4}\frac{-i}{k^2-i\epsilon} \int d^4x{\cal F}^{\mu \nu}(x){\cal F}_{\mu \nu}(x).
\label{lagunas}
\end{equation}
Here a cut-off  $\Lambda$ is introduced in the momentum integral:
\begin{equation}
 \int \frac{d^4k}{(2\pi)^4}\frac{-i}{k^2-i\epsilon} \simeq \frac{1}{16\pi^2}\Lambda^2
 \label{belenos}
 \end{equation}
 and the sign of the coefficient in (\ref{lagunas}) indicates at $D'<4$ asymptotic freedom. At $D'=4$ there is no effect.

In flat space and through use of direct momentum space integration without use of the proper time representation  the contributions in general gauges with gauge parameters $\omega_1, \omega_2$ can be arranged to cancel out.  It is first verified that (\ref{kildegaard}) and (\ref{kristofferryslinge}) cancel each other.  They are in flat space:
\begin{eqnarray}&&
i\kappa ^2\frac 12\frac {1}{\alpha ^2} \int d^4x \int d^4x'
<(\partial ^{\nu}h_{\mu \nu}
-\frac 12 \partial_{\mu}h^{\nu}\hspace{0.1 mm}_{\nu})(x)
(\partial^{\rho'}h_{\lambda'\rho'}-\frac 12 \partial _{\lambda'}h^{\rho'}\hspace{0.1 mm}_{\rho'})(x')>
\nonumber\\&&
<(\omega_1{\cal A}^{\mu}\partial _{\sigma}A^{\sigma}+\omega_2{\cal F}^{\sigma\mu}A_{\sigma})(x)(\omega_1{\cal A}^{\lambda'}\partial_{\gamma'}A^{\gamma'}+\omega_2{\cal F}^{\gamma'\lambda'}A_{\gamma'})(x')>
\nonumber\\&&
\label{stensvang}
\end{eqnarray}
and:
\begin{equation}
-\kappa^2\frac 12\frac {1}{\alpha}\int d^4xg^{\mu \nu}<(\omega_1 {\cal A}_\mu\partial_\lambda  A^\lambda +\omega_2{\cal F}_{\lambda \mu}A^\lambda)(x)(\omega_1{\cal A}_\nu \partial_\rho A^\rho +\omega _2{\cal F}_{\rho \nu}A^\rho)(x)>
\label{revninge}
\end{equation}
that cancel immediately by  application of (\ref{nissetid}).

While (\ref{revninge}) is a tadpole term, (\ref{stensvang}) in momentum space is a two-point self energy integral with the same structure as, for instance, the standard one-loop vacuum polarization integral of quantum electrodynamics.  For the latter, one has to chose a regularization scheme that keeps the vacuum polarization tensor transverse. This parallels the requirement that the regularization of (\ref{stensvang}) when converted to a momentum space integral should be carried out in such a way that (\ref{nissetid}) still applies, making the cancellation between (\ref{stensvang}) and (\ref{revninge}) possible.
This state of affairs should be compared to that found by proper time regularization, where the sum of (\ref{kildegaard}) and (\ref{kristofferryslinge}) is given by (\ref{birknakke}) containing quadratic and quartic divergences evaluated in Appendix A (the three expressions (\ref{birkverner}), (\ref{heroldsamson}) and (\ref{brotherhill})).

 In a similar way as (\ref{stensvang}) cancels with (\ref{revninge}) by momentum space integration, (\ref{primusbirk})   cancels with (\ref{edmundkristoffer}), when  converted to  momentum space integrals,  through the Ward identity (\ref{kattekilde}), which must be kept valid in the regularization procedure. The expressions corresponding to (\ref{borgmesteren}) and  (\ref{kristofferjuel}) vanish separately in momentum space.

Momentum space integration seems better off as a regularization procedure compatible with the Vilkovisky construction than the proper-time representation with a lower cut-off in the proper time integral. On the other hand, the proper-time representation allows a direct verification of the removal of the gauge parameter dependence from the finite and logarithmically dependent part of the effective action. Perhaps  a cut-off procedure could be found for the proper-time integrals that  is modeled after that of momentum space integration.

   \section{Conclusion}
   
   The following new results have been obtained in this article: The Vilkovisky construction was reconsidered and criteria for the applicability of a regularization scheme in this context were found. Also, the proper-time representation  of the effective action of one-loop quantum gravity was constructed for general gauges, the gauge parameter dependence was investigated, and it was found that the Vilkovisky construction removes  from it the finite and logarithmically divergent part but fails to do so from the quadratic and quartic divergences, and these conclusions were extended to the Maxwell-Einstein system. Using momentum-space integration in flat space instead it was found that these defects could be remedied for the Maxwell-Einstein system, suggesting that a modified cut-off procedure of the proper-time integrals should be chosen.
   
   \ \ \
 
 \noindent {\bf Acknowledgements:} I am grateful to Professor Francesco Sannino for organizing a meeting, which triggered this investigation, on the occasion of my retirement, and to  Professor Bo-Sture Skagerstam for a very inspiring correspondence. Also I  wish to thank Professor Sergei D. Odintsov for drawing my attention to \cite{Odintsov} and for sending me clarifying comments, and finally I would like to express my gratitude to two anonymous referees for constructive criticism of the first version of this paper.

\appendix

  \section{Heat kernel expansion}
   
  The ghost heat kernel defined by   (\ref{lindberg}) and (\ref{anna}) has the expansion \cite{DeWittII}:
    \begin{equation}
 h _{{\rm gh}\mu \xi'}(x, x'; \tau)=\frac{-i}{16\pi^2}\frac{1}{\tau^2}e^{i\frac{\sigma}{2\tau}}\Delta ^{\frac 12}\sum _{n=0}^\infty a_{n, {\rm gh}\mu \xi'}(x, x')(i\tau)^n
 \label{malacca}
 \end{equation}
where $\sigma$ is the geodesic interval between $x$ and $x'$ and $\Delta$ is the so-called Van Vleck determinant. At coinciding points one has:
  \begin{equation}
 a_{0, {\rm gh},  \mu \nu'}(x,x')\mid_{x'\rightarrow x}=g_{\mu \nu},, \ a_{1, {\rm gh},  \mu \nu'}(x,x')\mid_{x'\rightarrow x}\simeq -\frac R6g_{\mu \nu}-R_{\mu \nu}.
 \label{shorthorn}
 \end{equation}
 For the scalar heat kernel $h(x, x';\tau)$ defined by (\ref{lufthanna})  a corresponding expansion applies, with:
  \begin{eqnarray}&&
a_0(x, x')\mid_{x'\rightarrow x}=1, \  a_1(x, x')\mid_{x'\rightarrow x}  -\frac R6.
 \label{aurora}
 \end{eqnarray}
  Also:
 \begin{equation}
 \sigma_{;\lambda;\rho'}\simeq-g_{\lambda \rho'}
 \label{rudmeflux}
 \end{equation}
 for $x'\simeq x$. Hence it follows from  (\ref{malacca}):
 \begin{equation}
 h_{{\rm gh}\mu \xi'}(x, x';\tau)_{;\lambda;\rho'}=-g_{\lambda \rho'}\frac{1}{32\pi^2\tau^3}e^{i\frac{\sigma}{2\tau}}\Delta ^{\frac 12}\sum _{n=0}^\infty a_{n, {\rm gh}\mu \xi'}(x, x')(i\tau)^n+\cdots
 \label{malaya}
 \end{equation}
 where the remaining terms vanish at coinciding points.
 The graviton heat kernel $h^\alpha_{\mu \nu, \xi'\eta'}(x, x'; \tau)$ is by (\ref{lammelaar}) and (\ref{shorthorn}):
 \begin{equation}
 h^\alpha_{\mu \nu, \xi'\eta'}(x, x'; \tau)=\frac{-i}{16\pi^2}\frac{1}{\tau^2}e^{i\frac{\sigma}{2\tau}}\Delta ^{\frac 12}(\alpha(g_{\mu \xi'}g_{\nu \eta'}+g_{\mu \eta'}g_{\nu \xi'})-g_{\mu\nu}g_{\xi'\eta'})+\cdots.
 \label{cordova}
 \end{equation}

   The leading divergence of (\ref{mezquita})  at $\alpha =1$ is determined by the  quantity:
     \begin{eqnarray}&&
  \int d^4x'\int_0^1dt  
  h^\alpha_{\mu \nu,\xi'\eta'}(x,x';t\tau)\mid _{\alpha=1}(g^{\eta'\gamma'}{\cal F}^{\xi'\delta'}-\frac 14 g^{\xi'\eta'}{\cal F}^{\gamma'\delta'})(x')
  \nonumber\\&&
  (h^\beta _{\rho \delta'}(x,x';(1-t)\tau)_{;\lambda;\gamma'}
  -h^\beta _{\rho \gamma'}(x,x';(1-t)\tau)_{;\lambda;\delta'}-h^\beta _{\lambda \beta'}(x,x';(1-t)\tau)_{;\rho;\alpha'}
  \nonumber\\&&
  +h^\beta _{\lambda \gamma'}(x,x';(1-t)\tau)_{;\rho;\delta'})
     \nonumber\\&&
   \simeq -\frac{1}{16\pi^2\tau^3} \sqrt{-g} (g_{\mu \xi}g_{\nu \eta}+g_{\mu \eta}g_{\nu \xi}-g_{\mu \nu}g_{\xi\eta})(g^{\eta\gamma}{\cal F}^{\xi\delta}-\frac 14 g^{\xi\eta}{\cal F}^{\gamma\delta})(x)
   \nonumber\\&&
   (g_{\rho \delta}g_{\lambda \gamma}-g_{\lambda \delta}g_{\rho\gamma})
  \label{cornelia}
  \end{eqnarray}
  where the evaluation for simplicity can be carried out in flat space by Fourier transformation.
     In (\ref{belinda})   one encounters:
  \begin{eqnarray}&&
   \int d^4x\int _0^1 tdt h_{{\rm gh}\mu \xi'}(x, x';t\frac 1\alpha \tau) {\cal F}^{\mu \rho}(x)
 \nonumber\\&&
  (h^\beta _{\rho \delta'}(x,x'; (1-t)\tau)_{;\gamma';\eta'}
  -h ^\beta_{\rho \gamma'}(x,x'; (1-t)\tau)_{;\delta';\eta'})
   \nonumber\\&&
   \simeq \frac {\alpha^2}{4}\frac{1}{16\pi^2\tau^3}\sqrt{-g}(g_{\gamma'\eta'}{\cal F}_{\xi'\delta'}-g_{\delta'\eta'}{\cal F}_{\xi'\gamma'})(x')
\label{askemose}
\end{eqnarray}
  where the evaluation again most simply is carried out in flat space.

    The effective action in a general gauge in the heat-kernel representation also contains   the nonvanishing expressions (\ref{birknakke}),    (\ref{kristofferbroby}),   (\ref{borgmesteren}) and (\ref{kristofferjuel}).  
    They all contain a total derivative in the proper time integral and vanish in a formal sense in the same way as   (\ref{perritocaliente}). Nevertheless, they contain quartic or quadratic divergences. 
    Also the effective action contains   (\ref{juliechristie}) and (\ref{edmundkristoffer}), which cancel formally, but in fact have a quadratically divergent sum depending on the gauge parameters. The evaluation of  these quantities is sketched below; the calculation is most simply carried out in flat space and leads to the following intermediary results:  
     \begin{eqnarray}&&
   \int d^4x'\int_0^1 dt 
   h_{{\rm gh},\eta\mu'}(x, x';\frac 1\alpha t\tau){\cal F}^{\mu'\lambda'}(x')  h^\beta_{\lambda'\beta}(x', x;(1-t)\tau)
  \nonumber\\&&
   \simeq-\frac 14\alpha(3+\beta) \frac{i}{16\pi^2\tau^2}{\cal F}_{\eta\beta}(x)
  \label{birkikara}
  \end{eqnarray}
  and also:
   \begin{eqnarray}&&
\int d^4x '' \int_0^1 dth_{{\rm gh}, \mu\lambda''}(x, x'';\frac 1\alpha t\tau)
{\cal A}^{\lambda''}(x'')h(x'', x';\frac 1\beta(1-t)\tau)
 \nonumber\\&&
      \simeq -\alpha \beta\frac{i}{16\pi^2\tau^2} {\cal A}_{\mu}(x)
    \label{ascarithegreat}
\end{eqnarray}
that implies:
\begin{eqnarray}&&
\int d^4x '' \int_0^1 dth_{{\rm gh}, \mu\lambda''}(x, x'';\frac 1\alpha t\tau)
{\cal A}^{\lambda''}(x'')h(x'', x';\frac 1\beta(1-t)\tau)_{,\nu'}
 \nonumber\\&&
    \simeq -\frac 12 \alpha \beta\frac{i}{16\pi^2\tau^2} {\cal A}_{\mu, \nu'}(x).
    \label{nannafrisco}
\end{eqnarray}

    (\ref{birknakke}) is by  the Ward identity (\ref{monpetit}) the sum of three terms:
\begin{eqnarray}&&
-\frac 12\frac {\omega_2^2}{\alpha } \kappa ^2\int d^4x \int d^4x'
\int_0^\infty d\tau \frac{\partial}{\partial \tau}\tau \int_0^1 dth_{{\rm gh}, \lambda'\mu}(x', x;\frac 1\alpha t\tau)
\nonumber\\&&
{\cal F}^{\sigma\mu}(x){\cal F}^{\gamma'\lambda'}(x')h^{\beta}\hspace{0.1 mm}_{\sigma \gamma'} (x, x';(1-t)\tau)
\nonumber\\&&
\simeq -\frac 18(3+\beta)\omega_2^2\frac{i}{16\pi^2}\frac 1\tau\mid_{\tau\simeq 0}\kappa^2\int d^4x\sqrt{-g}{\cal F}^{\mu \nu}{\cal F}_{\mu \nu}(x)
\label{birkverner}
\end{eqnarray}
by (\ref{birkikara}), and also:
\begin{eqnarray}&&
\frac {\omega _1\omega_2}{\alpha } \kappa ^2\int d^4x \int d^4x'
\int_0^\infty d\tau \frac{\partial}{\partial \tau}\tau \int_0^1 dth_{{\rm gh}, \lambda'\mu}(x', x;\frac 1\alpha t\tau)
\nonumber\\&&
{\cal A}^{\mu}(x){\cal F}^{\gamma'\lambda'}(x')h(x, x';\frac 1\beta(1-t)\tau)_{;\gamma'}
\nonumber\\&&
\simeq  \frac 12\beta\omega _1\omega_2 \frac{i}{16\pi^2}\frac 1\tau\mid_{\tau\simeq 0}\kappa ^2\int d^4x\sqrt{-g}{\cal F}^{\mu\nu}{\cal A}_{\nu, \mu}(x)
\label{heroldsamson}
\end{eqnarray}
by (\ref{nannafrisco}), and finally:
\begin{eqnarray}&&
\frac 12\frac {\omega_1^2}{\alpha } \kappa ^2\int d^4x \int d^4x'
\int_0^\infty d\tau \frac{\partial}{\partial \tau}\tau \int_0^1 dth_{{\rm gh}, \lambda'\mu}(x', x;\frac 1\alpha t\tau)
\nonumber\\&&
{\cal A}^{\mu}(x){\cal A}^{\lambda'}(x')h(x, x';\frac 1\beta(1-t)\tau)_{;}\hspace{0.1 mm}^\sigma \hspace{0.1 mm}_{;\sigma}
\nonumber\\&&
= -i\frac 12\frac {\beta \omega_1^2}{\alpha } \kappa ^2\int d^4x \int d^4x'
\int_0^\infty d\tau \frac{\partial^2}{\partial \tau ^2}\tau \int_0^1 dth_{{\rm gh}, \lambda'\mu}(x', x;\frac 1\alpha t\tau)
\nonumber\\&&
{\cal A}^{\mu}(x){\cal A}^{\lambda'}(x')h(x, x';\frac 1\beta(1-t)\tau)
\nonumber\\&&
+i\frac 12\frac {\beta\omega_1^2}{\alpha } \kappa ^2\int d^4x\int_0^\infty d\tau \frac{\partial}{\partial \tau}h_{{\rm gh}, \mu\nu}(x, x;\frac 1\alpha \tau)
{\cal A}^{\mu }(x){\cal A}^{\nu}(x)
\nonumber\\&&
\simeq -\frac 12\beta(\alpha+\beta)\omega_1^2 \frac{1}{16\pi^2}\frac {1}{\tau^2}\mid_{\tau\simeq 0}\kappa^2\int d^4x\sqrt{-g}{\cal A}^\mu{\cal A}_\mu(x)
\label{brotherhill}
\end{eqnarray}
by (\ref{ascarithegreat}), where  only the quartic divergence was kept.

 The value of (\ref{kristofferbroby}) is by (\ref{birkikara}):
   \begin{eqnarray}&&
 \frac 14(3+\beta) \omega_2\frac{i}{16\pi^2}\frac 1\tau\mid_{\tau\simeq 0}\kappa^2\int d^4x\sqrt{-g}{\cal F}^{\mu \nu}{\cal F}_{\mu \nu}(x).
 \label{lewiskristoffer}
 \end{eqnarray}
 
 (\ref{borgmesteren}) contains two terms:
\begin{eqnarray}&&
\frac{\omega_1}{\sqrt{\alpha }}\kappa^2\int d^4x  {\cal A}^{\mu}(x)\int d^4x '\int_0^\infty d\tau\frac{\partial}{\partial \tau}\tau \int_0^1dt
h(x, x';\frac{t\tau}{\sqrt{\beta}})_{;}\hspace{0.1 mm}^{\nu'}
\nonumber\\&&
{\cal F}^{\lambda'}\hspace{0.1 mm}_{\nu'}(x')h_{{\rm gh}, \lambda'\mu}(x', x;\frac{(1-t)\tau}{\sqrt{\alpha}}))
\nonumber\\&&
\simeq -\frac 12 \sqrt{\beta}\omega_1 \frac{i}{16\pi^2}\frac 1\tau\mid_{\tau\simeq 0} \kappa^2\int d^4x\sqrt{-g}{\cal A}_{\nu, \mu}{\cal F}^{\mu\nu}
\label{kollekolle}
\end{eqnarray}
by (\ref{nannafrisco}), and:
\begin{eqnarray}&&
-\frac{\omega_1}{\sqrt \alpha}\kappa^2\int d^4x _1 {\cal A}^{\mu}(x)\int d^4x '\int_0^\infty d\tau\frac{\partial}{\partial \tau}\tau \int_0^1dt
\nonumber\\&&
h(x, x'; \frac{t\tau}{\sqrt \beta})_{;\nu';}\hspace{0.1 mm}^{\nu'}
{\cal A}^{\lambda '}(x')h_{{\rm gh}, \lambda'\mu}(x', x;\frac{(1-t)\tau}{\sqrt \alpha})
\nonumber\\&&
\simeq \sqrt{\beta(\alpha+\beta)}\omega_1\frac{1}{16\pi^2}\frac {1}{\tau^2}\mid_{\tau\simeq 0}\kappa^2\int d^4x\sqrt{-g}{\cal A}^\mu{\cal A}_\mu
\label{mckinney}
\end{eqnarray}
cp. (\ref{brotherhill}), where only the quartic divergence was determined.

(\ref{kristofferjuel}) is by (\ref{nannafrisco}):
\begin{equation}
-\frac 12\sqrt{\beta}\omega _2\frac{i}{16\pi^2} \frac 1\tau \mid _{\tau \simeq 0} \kappa^2\int d^4x\sqrt{-g}{\cal A}_{\nu, \mu}{\cal F}^{\mu\nu}.
\label{kingbunter}
\end{equation}

 We then consider (\ref{juliechristie}) and (\ref{edmundkristoffer}), using flat space heat kernels. The first term of (\ref{juliechristie}) has the quadratic divergence:
 \begin{equation}
\alpha \omega_2 \frac{i}{16\pi^2}\frac{1}{\tau }\mid _{\tau \simeq 0}
\kappa^2\int d^4x{\cal F}^{\mu \nu}{\cal F}_{\mu \nu}(x)  
 \label{jadekat}
 \end{equation}
 while the second term of (\ref{juliechristie})  is evaluated by means of (\ref{ascarithegreat}) which implies:
 \begin{eqnarray}&&
 \int d^4x '' \int_0^1 dth_{{\rm gh}, \mu\lambda''}(x, x'';\frac 1\alpha t\tau)
{\cal F}^{\lambda''\rho''}(x'')h(x'', x';\frac 1\beta(1-t)\tau)_{;\rho'';\sigma'}
 \nonumber\\&&
     \simeq -\frac 14\alpha \beta(\alpha +\beta)\frac{1}{16\pi^2\tau^3}{\cal F}_{\mu \sigma'}(x)
    \label{francesthegreat}
\end{eqnarray}
and so the second term of (\ref{juliechristie}) is:
 \begin{equation}
- \frac 14(\alpha+\beta)\omega_2\frac{i}{16\pi^2}\frac{1}{\tau }\mid _{\tau \simeq 0}\kappa^2\int d^4x{\cal F}^{\mu \nu}{\cal F}_{\mu \nu}(x).
 \label{tinkerogtanker}
 \end{equation}
The value of (\ref{edmundkristoffer}) is obtained from (\ref{jadekat}) and (\ref{tinkerogtanker}) by changing sign and replacing $\alpha$ by $\sqrt \alpha$ and $\beta$ by $\sqrt \beta$.
 
 \section{The Vilkovisky construction in next-lowest order}
 
 In the Landau-De Witt gauge we require that the gauge condition (\ref{basilosaurus}) is chosen
 such that only the Christoffel connection coupling term (\ref{wildfang})
  survives in (\ref{vildtfoged}).
 For the Maxwell-Einstein system the form of this gauge condition is:
 \begin{equation}
 \chi=\chi_\lambda=0
 \label{munbak}
 \end{equation}
  with:
  \begin{equation}
  \chi=-\sqrt{-g} A_{\mu;}\hspace{0.1 mm}^\mu
  \label{mari}
  \end{equation}
  and:
 \begin{eqnarray}&&
 \chi_\lambda=- \sqrt{-g}(h_{\lambda \mu;}\hspace{0.1 mm}^\mu-\frac 12 h_\mu\hspace{0.1 mm}^\mu\hspace{0.1 mm}_{;\lambda}+\kappa (A_{\mu ;}\hspace{0.1 mm}^{\mu}{\cal A}_\lambda+ {\cal F}_{\mu \lambda}A^\mu))
 \label{tyrannosaurus}
 \end{eqnarray}
 which is the gauge condition (\ref{samsonvullerslev})  with $\omega_1=\omega_2=1$ (the sign is unimportant). However, it was shown that the Vilkovisky construction of quantum gravity is sufficient formally (i.e. disregarding (\ref{bergerac}), (\ref{shehrezade}), (\ref{chabrol}), (\ref{shirvan})  and (\ref{perritocaliente})) to remove the dependence of the effective action on the dependence on the gauge parameter $\alpha$, and also that the additional terms of the effective action in gauges  with general values of the gauge parameters $\omega_1$ and $\omega_2$ formally cancel. One has to conclude that the Landau-DeWitt gauge appropriate for the Maxwell-Einstein system effective action taken to second order in $\kappa {\cal F}_{\mu \nu}$ is the same as that of quantum gravity, which is obtained by taking $\kappa \rightarrow 0$ in (\ref{tyrannosaurus}), while the Landau-DeWitt gauge condition obtained from (\ref{tyrannosaurus}) itself only is relevant to fourth order in $\kappa {\cal F}_{\mu \nu}$. That this is indeed the case follows from a detailed examination of the terms of (\ref{vildtfoged}) in this order of Maxwell-Einstein theory.

  The Ward identity corresponding to (\ref{royalacademy}) is first  determined.
 The graviton field two-point function  $<h_{\mu \nu}(x)h_{\xi'\eta'}(x')>$ is related to the heat kernel $h^\alpha_{\mu \nu, \xi'\eta'}(x,x';\tau)$ through (\ref{Magenta})
and the ghost two-point function $ <\xi_\mu(x)\bar \xi_\nu(y)>$ is expressed through the heat kernel $h_{{\rm gh}, \mu , \xi'}(x,x';\tau)$  in (\ref{askeladden}).
  (\ref{kalmar}) implies by   (\ref{Magenta}) and (\ref{askeladden})  the  Ward identity (\ref{royalacademy}) specialized to quantum gravity  and relating graviton and ghost two-point functions, cp. (\ref{kattekilde}): 
  \begin{eqnarray}&&
<(h_{\mu \nu;}\hspace{0.1 mm}^\mu-\frac 12 h_\mu \hspace{0.1 mm}^\mu \hspace{0.1 mm}_{;\nu})(x)h_{\lambda \rho}(y)>\simeq-\sqrt {\alpha}<(\xi_{\lambda;\rho}+\xi_{\rho;\lambda})(y)\bar \xi_\nu(x)>
\nonumber\\&&
+i\sqrt \alpha \int d^4w\sqrt{-g}<\xi ^\sigma (w)\bar \xi_\nu(x)>{\cal G}^{\omega\delta}(w)<(2h _{\omega \sigma;\delta}-h_{\omega \delta;\sigma})(w)h_{ \lambda\rho}(y)>.
\nonumber\\&&
   \label{Jekyll}
 \end{eqnarray}
  In the same way one gets from (\ref{ricciardelli}), using also (\ref{kalmar}) with the term involving the Einstein tensor disregarded:
 \begin{eqnarray}&&
 \alpha \frac{\partial}{\partial \alpha}<h_{\mu \nu}(x)h_{\lambda'\rho'}(x')>
 \nonumber\\&&
 \simeq i\int d^4x''\sqrt{-g''}<(\xi _{\mu;\nu}+\xi_{\nu;\mu})(x)\bar \xi ^{\sigma''}(x'')><(\xi_{\lambda';\rho'}+\xi_{\rho';\lambda'})(x')\bar \xi_{\sigma''}(x'')>.
 \nonumber\\&&
 \label{Violet's}
 \end{eqnarray}

  If the Ward identity (\ref{Jekyll}) is used with the second term on the right hand side included, and the replacement ${\cal G}^{\mu \nu}\rightarrow {\cal G}^{\mu \nu}-{\cal T}^{\mu \nu}$ next is  made, additional terms arise from (\ref{kildegaard}) and (\ref{primusbirk}):
\begin{eqnarray}&&
\frac 1\alpha \kappa^2\int d^4x_1\sqrt{-g_1} g^{\mu _1\nu_1} \int d^4x_2\sqrt{-g} g^{\mu _2\nu_2}\int d^4w\sqrt{-g}<\xi ^\sigma (w)\bar \xi_{\mu_1}(x_1)>
\nonumber\\&&
({\cal G}^{\omega\delta}-{\cal T}^{\omega \delta})(w)<(\xi  _{\sigma ;\omega; \delta}-R^\upsilon \hspace{0.1 mm}_{\omega \delta \sigma}\xi_\upsilon)(w)\bar \xi_{\mu _2}(x_2)>
  \nonumber\\&&
  <(\omega_1{\cal A}_{\nu_1}A^{\rho_1}\hspace{0.1 mm}_{;\rho_1}+\omega_2{\cal F}_{\rho _1\nu_1}A^{\rho_1})(x_1)(\omega_1{\cal A}_{\nu_2}A^{\rho_2}\hspace{0.1 mm}_{;\rho_2}+\omega_2{\cal F}_{\rho _2\nu_2}A^{\rho_2})(x_2)>
  \nonumber\\&&
 \label{palindrom}
\end{eqnarray}
and:
\begin{eqnarray}&&
\frac{1}{\sqrt \alpha}\kappa^2\omega_2\int d^4x_1\sqrt{-g_1}\int d^4x_2\sqrt{-g_2}\int d^4w\sqrt{-g}<\xi ^\sigma (w)\bar \xi_{\mu_1}(x)>
\nonumber\\&&
({\cal G}^{\omega\delta}-{\cal T}^{\omega \delta})(w)<(2h _{\omega \sigma;\delta}-h_{\omega \delta;\sigma})(w)h_{ \mu_2\nu_2}(x_2)>
\nonumber\\&&
g^{\mu_1\rho_1}{\cal F}_{\lambda_1\rho_1}(x_1)(g^{\mu_2\lambda_2}{\cal F}^{\nu_2\rho_2}-\frac 14 g ^{\mu_2\nu_2}{\cal F}^{\lambda _2\rho_2})(x_2)<A^{\lambda _1}(x_1)F_{\lambda_2\rho_2}(x_2)>.
\nonumber\\&&
\label{palimpsest}
\end{eqnarray}
These expressions are partially of fourth order in $\kappa$ and in the background field ${\cal A}_\mu$. The presence of these fourth order terms can be proven directly by a lengthy calculation.

$N^{\alpha \beta}$  now has components at first order in $\kappa $:
\begin{eqnarray}&&
N^{\xi_\alpha(x)c(y)} 
\nonumber\\&&
\simeq \kappa \frac{1}{\sqrt{\alpha \beta}}\int d^4w\sqrt{-g} <\xi_\alpha (x)((\bar \xi_\lambda{\cal A}^\lambda)_{;}\hspace{0.1 mm}^\mu+\bar \xi_\lambda(w){\cal F}^{\lambda\mu})(w)><c_{,\mu}(w)\bar c(y)>).
\nonumber\\&&
\label{rizzio}
\end{eqnarray}
At second order (\ref{lloydbridges}) is modified to:
\begin{eqnarray}&&
 N^{\xi_\mu(x) \xi_\nu(y)}\simeq 
   i\frac{1}{\sqrt \alpha}<\xi_\mu (x)\bar \xi _\nu(y)>
  +\kappa^2\frac{1}{\alpha}\int d^4w\int d^4z\sqrt{-g}<\xi_\mu(x)\bar \xi_\omega(w)>{\cal F}^{\omega \lambda}(w)
  \nonumber\\&&
  \Pi_{\lambda \rho} (w, z)  {\cal F}^{\sigma \rho}(z)< \xi_\sigma)(z)\bar \xi _\nu(y)>
 \label{makingwhoopee}
 \end{eqnarray}
 where $\Pi_{\lambda \rho}$ was introduced in (\ref{deichgraf}). 
  
  The projection operator (\ref{carolina}) is unmodified in lowest order but gets an additional term at second order in $\kappa$:
   \begin{eqnarray}&&
 \Delta  \Pi^{h_{\mu \nu}(x)}\hspace{0.1 mm}_{h_{\lambda \rho}(y)}
 \nonumber\\&&
= -\frac 1\alpha\kappa^2\hspace{0.1 mm}^4\sqrt{-g}\int d^4w\int d^4z<(\xi_{\mu;\nu}+\xi_{\nu;\mu})(x)\bar \xi^\gamma(w)>\hspace{0.1 mm}^4\sqrt{-g}{\cal F}_{\gamma \alpha}(w)
 \nonumber\\&&
 \Pi ^{\alpha\beta}(w, z)
{\cal F}_{\eta \beta}(z)\hspace{0.1 mm}^4\sqrt{-g} < \xi^\eta(z)\frac 12 (\bar \xi^ \lambda\hspace{0.1 mm}_{;}\hspace{0.1 mm}^\rho +\bar \xi^ \rho\hspace{0.1 mm}_{;}\hspace{0.1 mm}^\lambda-g^{\lambda \rho}\bar \xi_{\sigma;}\hspace{0.1 mm}^\sigma)(y)>^4\sqrt{-g}.
\nonumber\\&&
 \label{margolotta}
 \end{eqnarray} 
 A mixed  projection operator is by (\ref{obligation}) and (\ref{rizzio}):
 \begin{eqnarray}&&
  \Pi^{A_{\mu }(x)}\hspace{0.1 mm}_{h_{\lambda \rho}(y)}
 \nonumber\\&&
 =-i\frac {1}{\sqrt \alpha}\kappa \int d^4w\Pi _{\mu}\hspace{0.1 mm}^\nu  (x, w)
 ^4\sqrt{-g} {\cal F}_{\sigma \nu}(w)\frac{1}{\sqrt \alpha}<\xi^\sigma(w)
 \frac 12 (\bar \xi^ \lambda\hspace{0.1 mm}_{;}\hspace{0.1 mm}^\rho +\bar \xi^ \rho\hspace{0.1 mm}_{;}\hspace{0.1 mm}^\lambda-g^{\lambda \rho}\bar \xi_{\omega;}\hspace{0.1 mm}^\omega)(y)>^4\sqrt{-g}.
 \nonumber\\&&
 \label{uberwald}
 \end{eqnarray}

 In the Vilkovisky construction new terms of order $\kappa^2$ in (\ref{vildtfoged})  originate from:
  \begin{eqnarray}&&
\frac 12 \int d^4x \int d^4y \int d^4 z\int d^4 w\int d^4 u\int d^4 tS_{, h_{\mu\nu}(x)}R^{h_{\mu \nu}(x)}\hspace{0.1 mm}_{\xi_\omega(y), h_{\lambda \rho}(z)}N^{\xi_\omega(y)\xi_\sigma (w)}
  \nonumber\\&&
  R^{h_{\xi \eta}(u)}\hspace{0.1 mm}_{\xi_\sigma(w)}G_{h_{\xi \eta}(u)h_{ \alpha \beta}(t)}
 <h_{\alpha \beta}(t)h_{\lambda \rho}(z)>
  \nonumber\\&&
  +\frac 12 \int d^4x \int d^4y \int d^4 z\int d^4 w\int d^4 u\int d^4 tS_{, h_{\mu\nu}(x)}R^{h_{\mu \nu}(x)}\hspace{0.1 mm}_{\xi_\omega(y), h_{\lambda \rho}(z)}
 \nonumber\\&&
 (N^{\xi_\omega(y)\xi_\sigma (w)} R^{A_{\xi }(u)}\hspace{0.1 mm}_{\xi_\sigma(w)}+ N^{\xi_\omega(y)c (w)} R^{A_{\xi }(u)}\hspace{0.1 mm}_{c(w)})G_{A_{\xi}(u)A_{ \alpha }(t)}
 <A_{\alpha }(t)h_{\lambda \rho}(z)>^4\sqrt{-g}
 \nonumber\\&&
  \label{sophiereventlow}
  \end{eqnarray}
and:
 \begin{eqnarray}&&
 \frac 12\int d^4x \int d^4y \int d^4 z\int d^4 w\int d^4 u\int d^4 t\int d^4 rS_{, h_{\mu\nu}(x)}R^{h_{\mu \nu}(x)}\hspace{0.1 mm}_{\xi_\omega(y), h_{\lambda \rho}(z)}N^{\xi_\omega(y)\xi_\sigma (w)}
 \nonumber\\&&
 R^{h_{\xi \eta}(u)}\hspace{0.1 mm}_{\xi_\sigma(w)}G_{h_{\xi \eta}(u)h_{ \alpha \beta}(t)}
 <h_{\alpha \beta}(t)h_{\gamma \delta}(r)>^4\sqrt{-g}\Pi^{h_{\lambda \rho}(z)}\hspace{0.1 mm}_{h_{\gamma \delta}(r)}
 \nonumber\\&&
 +\frac 12  \int d^4x \int d^4y \int d^4 z\int d^4 w\int d^4 u\int d^4 t\int d^4 rS_{, h_{\mu\nu}(x)}R^{h_{\mu \nu}(x)}\hspace{0.1 mm}_{\xi_\omega(y), h_{\lambda \rho}(z)}N^{\xi_\omega(y)\xi_\sigma (w)}
 \nonumber\\&&
R^{h_{\xi \eta}(u)}\hspace{0.1 mm}_{\xi_\sigma(w)}G_{h_{\xi \eta}(u)h_{ \alpha \beta}(t)} <h_{\alpha \beta}(t)A_\gamma(r) >^4\sqrt{-g}\Pi^{h_{\lambda \rho}(z)}\hspace{0.1 mm}_{A_\gamma}(r)
 \nonumber\\&&
 + \frac 12\int d^4x \int d^4y \int d^4 z\int d^4 w\int d^4 u\int d^4 t\int d^4 rS_{, h_{\mu\nu}(x)}R^{h_{\mu \nu}(x)}\hspace{0.1 mm}_{\xi_\omega(y), h_{\lambda \rho}(z)}
 \nonumber\\&&
 (N^{\xi_\omega(y)\xi_\sigma (w)}
 R^{A_{\xi }(u)}\hspace{0.1 mm}_{\xi_\sigma(w)}+N^{\xi_\omega(y)c(w)}R^{A_{\xi }(u)}\hspace{0.1 mm}_{c(w)})G_{A_{\xi}(u)A_{ \alpha }(t)}
 <A_{\alpha }(t)h_{\gamma \delta}(r)>^4\sqrt{-g}\Pi^{h_{\lambda \rho}(z)}\hspace{0.1 mm}_{h_{\gamma \delta}(r)}
\nonumber\\&&
 + \frac 12\int d^4x \int d^4y \int d^4 z\int d^4 w\int d^4 u\int d^4 t\int d^4 rS_{, h_{\mu\nu}(x)}R^{h_{\mu \nu}(x)}\hspace{0.1 mm}_{\xi_\omega(y), g_{\lambda \rho}(z)}
 \nonumber\\&&
 (N^{\xi_\omega(y)\xi_\sigma (w)}
 R^{A_{\xi }(u)}\hspace{0.1 mm}_{\xi_\sigma(w)}+N^{\xi_\omega(y)c(w)}R^{A_{\xi }(u)}\hspace{0.1 mm}_{c(w)})G_{A_{\xi}(u)A_{ \alpha }(t)}
 <A_{\alpha }(t)A_{\gamma }(r)>^4\sqrt{-g}\Pi^{h_{\lambda \rho}(z)}\hspace{0.1 mm}_{A_{\gamma }(r)}.
 \nonumber\\&&
  \label{sophielevetzau}
 \end{eqnarray}
   These expressions are now analyzed and shown to cancel  with (\ref{palindrom}) and (\ref{palimpsest}); also it is shown that they vanish, if the Landau-DeWitt gauge conditions (\ref{munbak}) are imposed. 
   
   The parts of (\ref{sophiereventlow}) and (\ref{sophielevetzau}) involving the graviton correlation function \\$<h_{\alpha \beta}(t)h_{\gamma \delta}(r)>$  at second order in $\kappa$ constructed by means of the couplings (\ref{anaximander}) and (\ref{mattgroening})  involves two factors $S_{, h_{\mu\nu}}$; the argument amounts to using (\ref{Jekyll})  with the replacement ${\cal G}^{\mu \nu}\rightarrow {\cal G}^{\mu \nu}-{\cal T}^{\mu \nu}$ in the second term on the right hand side. Consequently these terms are disregarded in the approximation where only one derivative of the classical action is kept in (\ref{vildtfoged}).

   There is  an extra term in (\ref{sophiereventlow}) from the second order term of (\ref{makingwhoopee}):
 \begin{eqnarray}&&
  \frac 12\frac 1\alpha\kappa^2\int d^4x\sqrt{-g}\int d^4y\int d^4w\int d^4z\sqrt{-g}\sqrt{-g}({\cal G}^{\mu\nu}-{\cal T}^{\mu \nu})(x)
 \nonumber\\&&
<(h_{\omega \tau;}\hspace{0.1 mm}^\tau -\frac 12h_\tau \hspace{0.1 mm}^\tau \hspace{0.1 mm}_{;\omega})(y)(h_{\mu\lambda}\xi^\lambda\hspace{0.1 mm}_{;\nu}+h_{\nu\lambda}\xi^\lambda  \hspace{0.1 mm}_{;\mu}+\xi^\lambda h_{\mu\nu;\lambda})(x)\bar \xi_\sigma(w)>{\cal F}^{\sigma \lambda}(w)
  \nonumber\\&&
  \Pi_{\lambda \rho} (w, z)  {\cal F}^{\sigma \rho}(z)< \xi_\sigma(z)\bar \xi^\omega(y)>
 \label{benghazi}
 \end{eqnarray}
 that vanishes in the Landau-DeWitt gauge to order $\kappa^2$ since the additional term in (\ref{tyrannosaurus}) makes the expression $O(\kappa^3)$.  The corresponding term of (\ref{sophielevetzau}) vanishes.

   In (\ref{sophiereventlow}) and (\ref{sophielevetzau})  the following combination is present:
   \begin{eqnarray}&&
   \int d^4w\int d^4u(N^{\xi_\omega(y)\xi_\sigma (w)} R^{A_{\xi }(u)}\hspace{0.1 mm}_{\xi_\sigma(w)}+N^{\xi_\omega(y)c (w)} R^{A_{\xi }(u)}\hspace{0.1 mm}_{c(w)})G_{A_{\xi}(u)A_{ \alpha }(t)}
A_{\alpha }(t))
\nonumber\\&&
=-i \frac{1}{\sqrt \alpha}\kappa\int d^4wA^\lambda (x)\Pi_{\lambda \mu}(x, w){\cal F}^{\mu \rho}(w)<\xi_\rho(w)\bar\xi_\omega(y)>
   \label{odingaard}
   \end{eqnarray}
    and thus one gets by (\ref{uberwald}):
\begin{eqnarray}&&
 \frac 12\int d^4x \int d^4y \int d^4 z\int d^4 w\int d^4 u\int d^4 t\int d^4 rS_{, h_{\mu\nu}(x)}R^{h_{\mu \nu}(x)}\hspace{0.1 mm}_{\xi_\omega(y), g_{\lambda \rho}(z)}
 \nonumber\\&&
 (N^{\xi_\omega(y)\xi_\sigma (w)}
 R^{A_{\xi }(u)}\hspace{0.1 mm}_{\xi_\sigma(w)}+N^{\xi_\omega(y)c(w)}R^{A_{\xi }(u)}\hspace{0.1 mm}_{c(w)})G_{A_{\xi}(u)A_{ \alpha }(t)}
 <A_{\alpha }(t)A_{\gamma }(r)>^4\sqrt{-g}\Pi^{h_{\lambda \rho}(z)}\hspace{0.1 mm}_{A_{\gamma }(r)}
 \nonumber\\&&
\simeq
\frac 1\alpha \kappa^2 \int d^4x\sqrt{-g}\int d^4y\int d^4w\int d^4t({\cal G}^{\mu\nu}(x)-{\cal T}^{\mu \nu})(x)
\nonumber\\&&
{\cal F}^{\omega \rho}(w)^4\sqrt{-g}\Pi_{\omega \lambda}(w,t)^4\sqrt{-g}<A^\lambda(t)A_\gamma(r)>{\cal F}^{\sigma \gamma}(r)
\nonumber\\&&
(<\xi_{\upsilon;\mu}(x)\bar \xi_\sigma(w)><\xi ^\upsilon \hspace{0.1 mm}_{;\nu}(x)\bar \xi _\rho(y)>
-R_{\upsilon \mu \phi \nu}(x)<\xi^\upsilon(x)\bar\xi_\sigma(w)><\xi^\phi (x)\bar \xi _\rho(y)>).
\label{rabsenfuchs}
\end{eqnarray}

 In (\ref{sophielevetzau}) one gets by    (\ref{margolotta}):
  \begin{eqnarray}&&
   \int d^4w\int d^4u \int d^4tR^{g_{\xi \eta}(w)}\hspace{0.1 mm}_{\xi_\mu(x)}G_{\xi \eta, \alpha \beta}(w, u)
 ^4\sqrt{-g}<h_{\alpha \beta}(u)h_{\sigma \omega }(t)>^4\sqrt{-g}\Pi^{g_{\lambda \rho}(z)}\hspace{0.1 mm}_{g_{\sigma \omega}(t)}
 \nonumber\\&&
 \rightarrow i\kappa^2\hspace{0.1 mm}^4\sqrt{-g}\int d^4w\int d^4z\sqrt{-g}\int d^4u\sqrt{-g} {\cal F} ^{\omega \sigma}(w)<\xi_{\omega}(w)\bar \xi_{ \mu}(x)>\Pi_{\sigma \gamma}(w, u)^4\sqrt{-g}{\cal F}^{\beta\gamma }(u)
  \nonumber\\&&
  <\xi _\beta (u)(\bar \xi_{\lambda;\rho}+\bar \xi_{\rho;\lambda})(z)>.
    \label{scheherezade}
  \end{eqnarray}
   Also one finds in (\ref{sophielevetzau}) by (\ref{Jekyll}):
   \begin{eqnarray}&&
 \int d^4w\int d^4u \int d^4tR^{g_{\xi \eta}(w)}\hspace{0.1 mm}_{\xi_\mu(x)}G_{\xi \eta, \alpha \beta}(w, u)
 ^4\sqrt{-g}<h_{\alpha \beta}(u)A_{\gamma }(t)>^4\sqrt{-g}\Pi^{g_{\lambda \rho}(z)}\hspace{0.1 mm}_{A_{\gamma }(t)}
 \nonumber\\&&
\rightarrow  - i \kappa^2\hspace{0.1 mm}^4\sqrt{-g}\int d^4w\hspace{0.1 mm}^4\sqrt{-g}\int d^4u\hspace{0.1 mm}^4\sqrt{-g}{\cal F} ^{\omega \sigma}(w)<\xi_{\omega}(w)\bar \xi_{ \mu}(x)>\Pi_{\sigma \gamma}(w, u)^4\sqrt{-g}{\cal F}^{\beta\gamma }(u)
  \nonumber\\&&
  <\xi _\beta (u)(\bar \xi_{\lambda;\rho}+\bar \xi_{\rho;\lambda})(z)>
 \label{rystespyd}
 \end{eqnarray}
  where the correlation function $<A_\mu(x)h_{\lambda \rho}(z)>$ was formed by means of the coupling (\ref{anaximander}) with the splitting (\ref{stocksplit}), 
  and (\ref{rystespyd}) cancels with (\ref{scheherezade}).

 Other higher order terms constructed only by means of the coupling (\ref{anaximander}) are next considered.
 Using (\ref{odingaard})  one gets by (\ref{supergauge}): 
      \begin{eqnarray}&&
  \int d^4y \int d^4 z\int d^4 w\int d^4 u\int d^4 tR^{h_{\mu \nu}(x)}\hspace{0.1 mm}_{\xi_\omega(y), h_{\lambda \rho}(z)}
  \nonumber\\&&
  (N^{\xi_\omega(y)\xi_\sigma (w)} R^{A_{\xi }(u)}\hspace{0.1 mm}_{\xi_\sigma(w)}+N^{\xi_\omega(y)c (w)} R^{A_{\xi }(u)}\hspace{0.1 mm}_{c(w)})G_{A_{\xi}(u)A_{ \alpha }(t)}
<A_{\alpha }(t)h_{\lambda \rho}(z)>)
 \nonumber\\&&
 \rightarrow \frac{1}{\sqrt \alpha}\kappa^3\int d^4y\int d^4w \int d^4u{\cal F}_{\sigma \omega}(y)\hspace{0.1 mm}^4\sqrt{-g}\Pi ^{\sigma \tau}(y, w)\hspace{0.1 mm}^4\sqrt{-g}  <A_\tau(w)F_{\xi \eta}(u)>
 \nonumber\\&&
 (g^{\alpha \xi}{\cal F} ^{\beta \eta}-\frac 14g^{\alpha \beta}{\cal F}^{\xi \eta})(u)<h_{\alpha \beta}(u)(h_{\mu\lambda}\xi^\lambda\hspace{0.1 mm}_{;\nu}+h_{\nu\lambda}\xi^\lambda \hspace{0.1 mm}_{;\mu}+\xi^\lambda h_{\mu\nu;\lambda})(x)\bar \xi^\omega(y)>.
 \nonumber\\&&
 \label{mustang}
 \end{eqnarray}
  (\ref{mustang}) contributes to the effective action through (\ref{sophiereventlow}):
    \begin{eqnarray}&& 
 - \frac{1}{\sqrt \alpha}\kappa^2\int d^4x\sqrt{-g}\int d^4y\int d^4w\int d^4u\sqrt{-g}({\cal G}^{\mu\nu}-{\cal T}^{\mu \nu})(x)
 \nonumber\\&&
 {\cal F}_{\sigma \omega}(y)\hspace{0.1 mm}^4\sqrt{-g}\Pi ^{\sigma \tau}(y, w)\hspace{0.1 mm}^4\sqrt{-g}  <A_\tau(w)F_{\xi \eta}(u)>
 \nonumber\\&&
 (g^{\alpha \xi}{\cal F} ^{\beta \eta}-\frac 14g^{\alpha \beta}{\cal F}^{\xi \eta})(u)<h_{\alpha \beta}(u)(h_{\mu\lambda}\xi^\lambda\hspace{0.1 mm}_{;\nu}+\frac 12\xi^\lambda h_{\mu\nu;\lambda})(x)\bar \xi^\omega(y)>
\nonumber\\&&
 \label{richardgere}
 \end{eqnarray}
with the gauge dependent part  by   (\ref{Violet's}) and  (\ref{digegreve}):
  \begin{eqnarray}&& 
 - \frac 12\frac 1\alpha\kappa^2\int d^4x\sqrt{-g}\int d^4y\int d^4w\int d^4z\sqrt{-g}\sqrt{-g}({\cal G}^{\mu\nu}-{\cal T}^{\mu \nu})(x)
 \nonumber\\&&
<(h_{\omega \tau;}\hspace{0.1 mm}^\tau -\frac 12h_\tau \hspace{0.1 mm}^\tau \hspace{0.1 mm}_{;\omega})(y)(h_{\mu\lambda}\xi^\lambda\hspace{0.1 mm}_{;\nu}+h_{\nu\lambda}\xi^\lambda  \hspace{0.1 mm}_{;\mu}+\xi^\lambda h_{\mu\nu;\lambda})(x)\bar \xi_\sigma(w)>{\cal F}^{\sigma \lambda}(w)
  \nonumber\\&&
  \Pi_{\lambda \rho} (w, z)  {\cal F}^{\sigma \rho}(z)< \xi_\sigma(z)\bar \xi^\omega(y)>
   \label{captainlloyd}
 \end{eqnarray}
 that cancels with (\ref{benghazi}). 
  Also (\ref{sophielevetzau}) contains, cp. (\ref{richardgere}):
 \begin{eqnarray}&&
   - \frac{1}{\sqrt \alpha}\kappa^2\int d^4x\sqrt{-g}\int d^4y\int d^4w\int d^4u\int d^4t\sqrt{-g}({\cal G}^{\mu\nu}-{\cal T}^{\mu \nu})(x)
 \nonumber\\&&
 {\cal F}_{\sigma \omega}(y)\hspace{0.1 mm}^4\sqrt{-g}\Pi ^{\sigma \tau}(y, w)\hspace{0.1 mm}^4\sqrt{-g}  <A_\tau(w)F_{\xi \eta}(u)>
 \nonumber\\&&
 (g^{\alpha \xi}{\cal F} ^{\beta \eta}-\frac 14g^{\alpha \beta}{\cal F}^{\xi \eta})(u)\Pi ^{g_{\alpha \beta}(u)}\hspace{0.1 mm}_{g_{\gamma \delta }(t)}<h_{\gamma \delta}(t)(h_{\mu\lambda}\xi^\lambda\hspace{0.1 mm}_{;\nu}+\frac 12\xi^\lambda h_{\mu\nu;\lambda})(x)\bar \xi^\omega(y)>
 \nonumber\\&&
 \label{powershopping}
  \end{eqnarray}
with no gauge parameter dependence, keeping in mind that the normalization of the ghost propagator involves the gauge parameter $\alpha$.

 All dependence on the gauge parameters $\alpha$ and $\beta$ cancels out so far.
We then turn to terms constructed also from the coupling (\ref{tjalfekristoffer}).

  With two couplings (\ref{tjalfekristoffer}) one gets:
   \begin{eqnarray}&&
 \int d^4w\int d^4u R^{g_{\xi \eta}(w)}\hspace{0.1 mm}_{\xi_\mu(x)}G_{\xi \eta, \alpha \beta}(w, u)
<h_{\alpha \beta}(u)h_{\gamma \delta}(y)>
 \nonumber\\&&
 \rightarrow i\frac{1}{\sqrt \alpha}\kappa^2\int d^4w\sqrt{-g}<(\omega_1 {\cal A}_\mu A^\kappa \hspace{0.1 mm}_{;\kappa}+\omega _2{\cal F}_{\kappa \mu}A^\kappa)(x)
(\omega_1 {\cal A}^\rho A^\epsilon \hspace{0.1 mm}_{;\epsilon}+\omega _2{\cal F}^{\epsilon \rho}A_\epsilon)(w)>
\nonumber\\&&
<(\xi_{\gamma;\delta}+\xi_{\delta;\gamma})(y)\bar \xi_\rho (w)>
 \label{kasperroeghat}
 \end{eqnarray}
which does not contribute to (\ref{sophielevetzau}), while its contribution to  (\ref{sophiereventlow}) is:
     \begin{eqnarray}&&
   \frac 1\alpha\kappa^2\int d^4x \sqrt{-g}({\cal G}^{\mu\nu}-{\cal T}^{\mu \nu})(x)
 \nonumber\\&&
 \int d^4y\sqrt{-g}\int d^4w\sqrt{-g}<(\omega_1 {\cal A}_\tau A^\kappa \hspace{0.1 mm}_{;\kappa}+\omega _2{\cal F}_{\kappa \tau}A^\kappa)(y)
(\omega_1 {\cal A}^\rho A^\epsilon \hspace{0.1 mm}_{;\epsilon}+\omega _2{\cal F}^{\epsilon \rho}A_\epsilon)(w)>
\nonumber\\&&
(<\xi_{\lambda;\mu}(x)\bar \xi_\rho(w)><\xi ^\lambda \hspace{0.1 mm}_{;\nu}(x)\bar \xi ^\tau(y)>
-R_{\sigma \mu \lambda \nu}(x)<\xi^\sigma(x)\bar\xi_\rho(w)><\xi^\lambda (x)\bar \xi ^\tau(y)>)
\nonumber\\&&
\label{menhammar}
\end{eqnarray}
and (\ref{menhammar}) cancels with (\ref{palindrom}).  Also (\ref{menhammar}) cancels with (\ref{rabsenfuchs}) for $\omega_2=1$ and a transverse photon propagator (Landau-DeWitt gauge).

By (\ref{supergauge}) and  (\ref{odingaard}) one finds, forming the two-point correlation function $<A_{\alpha}(t)h_{\lambda \rho}(z)>$ by means of the coupling (\ref{tjalfekristoffer}):
 \begin{eqnarray}&&
  \int d^4y \int d^4 z\int d^4 w\int d^4 u\int d^4 tR^{g_{\mu \nu}(x)}\hspace{0.1 mm}_{\xi_\omega(y), g_{\lambda \rho}(z)}
  \nonumber\\&&
 (N^{\xi_\omega(y)\xi_\sigma (w)} R^{A_{\xi }(u)}\hspace{0.1 mm}_{\xi_\sigma(w)}G_{\xi, \alpha }(u,t)
 <A_{\alpha}(t)h_{\lambda \rho}(z)>
 \nonumber\\&&
 +N^{\xi_\omega(y)c (w)} R^{A_{\xi }(u)}\hspace{0.1 mm}_{c(w)}G_{\xi, \alpha }(u,t)
 <A_{\alpha}(t)h_{\lambda \rho}(z)>)
 \nonumber\\&&
 \rightarrow \frac{\omega_2}{\alpha}\kappa^4\int d^4y\sqrt{-g} \int d^4 w\int d^4u\sqrt{-g}{\cal F}_{\omega \sigma }(y)\hspace{0.1 mm}^4\sqrt{-g}\Pi^{\omega\tau}(y, w)\hspace{0.1 mm}^4\sqrt{-g} < A_\tau(w)A_\kappa(u)>{\cal F}^{\kappa \rho}(u)
 \nonumber\\&&
(<(\xi_{\mu;\lambda}+\xi_{\lambda;\mu})(x)\bar \xi_\rho(u)><\xi ^\lambda \hspace{0.1 mm}_{;\nu}(x)\bar \xi ^\sigma(y)>
+<(\xi_{\nu;\lambda}+\xi_{\lambda;\nu})(x)\bar \xi_\rho(u)><\xi ^\lambda \hspace{0.1 mm}_{;\mu}(x)\bar \xi ^\sigma(y)>
\nonumber\\&&
+<(\xi_{\mu;\nu;\lambda}+\xi_{\nu;\mu;\lambda})(x)\bar\xi_\rho(u)><\xi^\lambda (x)\bar \xi ^\sigma(y)>)
 \label{rabalder}
 \end{eqnarray}
contributing to (\ref{sophiereventlow}) and the trivial part of (\ref{sophielevetzau}), where only the term $\delta ^{(\mu \nu)}\hspace{0.1 mm}_{(\lambda \rho)}$ of the projection operator $\Pi^{h_{\lambda \rho}(y)}\hspace{0.1 mm}_{h_{\mu \nu}(x)}
$ is kept:
   \begin{eqnarray}&&
    -\frac 2\alpha\omega_2\kappa^2\int d^4x \sqrt{-g}({\cal G}^{\mu\nu}-{\cal T}^{\mu \nu})(x)
 \nonumber\\&&
\int d^4y\sqrt{-g} \int d^4 w\int d^4u\sqrt{-g}{\cal F}_{\omega \sigma}(y)\hspace{0.1 mm}^4\sqrt{-g}\Pi^{\omega\tau}(y, w)\hspace{0.1 mm}^4\sqrt{-g} < A_\tau(w)A_\kappa(u)>{\cal F}^{\kappa \rho}(u)
 \nonumber\\&&
(<\xi_{\lambda;\mu}(x)\bar \xi_\rho(u)><\xi ^\lambda \hspace{0.1 mm}_{;\nu}(x)\bar \xi ^\sigma(y)>
-R^\sigma\hspace{0.1 mm}_{\mu \lambda \nu}(x)<\xi_{\sigma}(x)\bar\xi_\rho(u)><\xi^\lambda (x)\bar \xi ^\sigma(y)>).
\nonumber\\&&
 \label{champignon}
 \end{eqnarray}

    Also using both the couplings (\ref{anaximander}) and (\ref{tjalfekristoffer}) to form a graviton two-point correlation function of second order in $\kappa$ one gets:
   \begin{eqnarray}&&
 \int d^4w\int d^4u R^{g_{\xi \eta}(w)}\hspace{0.1 mm}_{\xi_\mu(x)}G_{\xi \eta, \alpha \beta}(w, u)
<h_{\alpha \beta}(u)h_{\gamma \delta}(y)>
 \nonumber\\&&
 \rightarrow i\omega _2\kappa^2\hspace{0.1 mm}^4\sqrt{-g} \int d^4w\int d^4z<\xi_\lambda (w)\bar \xi_\mu(x)>{\cal F}^{\lambda \omega}(w)\hspace{0.1 mm}^4\sqrt{-g}\Pi _{\omega\delta}(w, z)\hspace{0.1 mm}^4\sqrt{-g}{\cal F}^{\delta\gamma  }(z)
\nonumber\\&&
<(\xi_{\alpha;\beta}+\xi_{\beta;\alpha})(y)\bar \xi_\gamma(z)>
\nonumber\\&&
+i\frac{\omega _2}{\sqrt \alpha}\kappa^2\sqrt{-g} \int d^4w\sqrt{-g}<h_{\alpha \beta}(y) h_{\lambda \rho}(w)>({\cal F}^{\lambda \omega}g^{\rho \sigma}-\frac 14g^{\lambda \rho}{\cal F}^{\sigma \omega})(w)<F_{\sigma \omega}(w)A^\upsilon(x)>{\cal F}_{\upsilon\mu}(x)
\nonumber\\&&
 \label{kratholm}
 \end{eqnarray}
  contributing to (\ref{sophiereventlow}) and the trivial part of (\ref{sophielevetzau}) first:
     \begin{eqnarray}&&
\frac{ 2 \omega _2}{\sqrt \alpha}\kappa^2\int d^4x\sqrt{-g}({\cal G}^{\mu\nu}-{\cal T}^{\mu \nu})(x)
 \nonumber\\&&
 \int d^4w\sqrt{-g}< h_{\tau \rho}(w)(h_{\mu\lambda}\xi^\lambda\hspace{0.1 mm}_{;\nu}+\frac 12\xi^\lambda h_{\mu\nu;\lambda})(x)\bar \xi^\kappa(y>
 \nonumber\\&&
 ({\cal F}^{\tau \omega}g^{\rho \sigma}-\frac 14g^{\tau \rho}{\cal F}^{\sigma \omega})(w)<F_{\sigma \omega}(w)A^\upsilon(x)>{\cal F}_{\upsilon\mu}(x)
  \label{bluebayou}
  \end{eqnarray}
  canceling with (\ref{palimpsest}).   For $\omega_2=1$ and a transverse graviton propagator (Landau-DeWitt gauge) (\ref{bluebayou})    cancels with (\ref{richardgere}) and (\ref{powershopping}).
   One also gets from  (\ref{kratholm}) the following contribution  to (\ref{sophiereventlow}) and the trivial part of (\ref{sophielevetzau}):
   \begin{eqnarray}&&
   -2\omega_2\kappa^2\int d^4x \sqrt{-g}({\cal G}^{\mu\nu}-{\cal T}^{\mu \nu})(x)
 \nonumber\\&&
  \int d^4w\int d^4z<\xi_\lambda (w)\bar \xi_\mu(x)>{\cal F}^{\lambda \sigma}(w)\hspace{0.1 mm}^4\sqrt{-g}\Pi _{\sigma \omega}(w, z)\hspace{0.1 mm}^4\sqrt{-g}{\cal F}^{\kappa \omega }(z)
\nonumber\\&&
 (<\xi_{\lambda;\mu}(x)\bar \xi_\rho(w)><\xi ^\lambda \hspace{0.1 mm}_{;\nu}(x)\bar \xi ^\tau(y)>
-R^\sigma \hspace{0.1 mm}_{\mu\lambda \nu}(x)<\xi_{\sigma}(x)\bar\xi_\rho(w)><\xi^\lambda (x)\bar \xi ^\tau(y)>).
\nonumber\\&&
  \label{tishemingo}
  \end{eqnarray}

 (\ref{sophielevetzau}) also contains  contributions  from the nontrivial part of the projection operator (\ref{carolina}) and from (\ref{uberwald}) and involving the coupling (\ref{tjalfekristoffer}) once.  After some calculation one finds that they cancel with (\ref{champignon}) and (\ref{tishemingo}).

 To summarize, a proof on the formal (non-regularized) level  has  been sketched in this appendix that the Vilkovisky construction given by (\ref{vildtfoged}) makes the effective action of the Maxwell-Einstein system gauge parameter independent at the one-loop level and at next-lowest order in the gravitational coupling constant $\kappa$. In the course of the proof it was found that the Landau-DeWitt gauge given by (\ref{munbak}), (\ref{mari}) and  (\ref{tyrannosaurus}) and with also the $O(\kappa)$ term included in (\ref{tyrannosaurus}) makes the expressions (\ref{sophiereventlow}) and (\ref{sophielevetzau}) vanish, i.e. both the terms of (\ref{vildtfoged}) not involving the field space connection, in contrast to the lowest order calculation, where  the   terms  of the gauge condition (\ref{tyrannosaurus})  involving only the field $h_{\mu\nu}$  are sufficient for this effect.

\end{document}